# Evidence of new twinning modes in magnesium questioning the shear paradigm

C. Cayron[1], R.E. Logé[1]

**Twinning is an important deformation mode of hexagonal close-packed metals. The crystallographic theory is based on the 150-years old concept of simple shear. The habit plane of the twin is the shear plane; it is invariant. Here we present Electron BackScatter Diffraction observations and crystallographic analysis of a millimeter size twin in a magnesium single crystal whose straight habit plane, unambiguously determined both the parent crystal and in its twin, is not an invariant plane. This experimental evidence demonstrates that macroscopic deformation twinning can be obtained by a mechanism that is not a simple shear. Beside, this unconventional twin is often co-formed with a new conventional twin that exhibits the lowest shear magnitude ever reported in metals. The existence of unconventional twinning introduces a shift of paradigm and calls for the development of a new theory for the displacive transformations.**

Deformation twinning is an important deformation mode in hexagonal-close packed (hcp) materials, such as titanium, zirconium and magnesium alloys and is the subject of active research. The considered general theory is 150 years old and is based on the concept of simple shear. In 1867, Thomson and Guthrie Tait[1] defined a simple shear as *"the property that two kinds of planes (two different sets of parallel planes) remain unaltered, each in itself"*, which lead to the nomenclature ($K_1,\eta_1,K_2,\eta_2$) introduced in 1889 by Mügge[2,3]. The first plane is the shear plane $K_1$; it is untilted and undistorted; it contains the shear direction $\eta_1$. The second plane $K_2$ is undistorted but rotated. The direction $\eta_2$ belongs to $K_2$ and is perpendicular to the intersection between $K_1$ and $K_2$. Between 1950 and 1970, the theory of twinning was mathematically developed and refined with linear algebra[4-9]. The theory uses three matrices: the simple shear matrix **S** deduced from ($K_1,\eta_1$), the correspondence matrix **C**, and the parent/twin misorientation matrix **T**. More details on these matrices are given in the preliminary section of **Supplementary Equations**. Among the numerous but finite possible correspondence and shear matrices resulting from the calculations, only those with the lowest shear magnitude are considered as realistic. As an instantaneous and homogeneous shear is not possible at reasonable stresses[10], the simple shear was assumed to be produced by the coordinate propagation of "twinning dislocations" created by complex "pole mechanisms"[11-13]. The generalization of dislocations as a fundamental part of the transformation mechanism really came

---

[1] Laboratory of ThermoMechanical Metallurgy (LMTM), PX Group Chair, Ecole Polytechnique Fédérale de Lausanne (EPFL), Rue de la Maladière 71b, 2000 Neuchâtel, Switzerland

with the introduction of the "disconnections"[14,15]. All these cited models are based on the shear paradigm; they have dominated the theoretical developments of deformation twinning over the last seventy years.

The formation of $\{10\bar{1}2\}$ extension twins without straight shear plane was recently observed by in-situ Transmission Electron Microscopy (TEM) in magnesium nano-pillars[16]. The twins are characterized by a parent/twin misorientation of 90° around the **a**-axis, instead of 86° for the conventional extension twins in bulk magnesium. These observations, and earlier molecular-dynamic simulations of the nucleation stage of extension twinning[17,18], lead some researchers to propose a new twinning mechanism based on "*pure shuffle*", or equivalently "*zero shear*"[16,19]. This mechanism is the subject of an intense debate[20]. One way to reduce the controversy was to admit that the "zero shear" mechanism "*distinctively differs from any other twinning modes*" and that "*this should not be deemed as the failure of the classical theory*"[19]. Is that correct? Is the unconventional (90°, **a**) twin an exotic case limited only to extension twinning in hcp metals, and even more specifically to the nucleation step or to nano-sized samples? It was recently shown that the unconventional "zero-shear" (90°, **a**) and conventional shear (86°, **a**) twins actually result from the same distortion because they differ just only by an obliquity correction[21]. The model assumes that the atoms move as hard-spheres, and it calculates for a given orientation relationship the analytical forms of the atomic trajectories, lattice distortion, and volume change. A similar approach was used to model $\{10\bar{1}1\}$ contraction twinning[22]. The volume change is a direct consequence of the hard-sphere assumption. Indeed, the Kepler conjecture (demonstrated by Hales[23]) implies that all the intermediate states between an hcp structure and its twin have a density lower than that of hcp. The volume change is not negligible; for magnesium, it is 3% for extension twinning[21] and 5% for contraction twinning[22]. The same approach was used for martensitic transformations between fcc, bcc and hcp phases[24]. It should be noted that a volume change is not compatible with a simple shear. Beside the volume change, the calculations proved that the habit plane is not invariant; it is untilted but distorted, and restored only when the process is complete. Thus, one can ask whether for some twins the interface plane could be transformed into another crystallographic plane. This would then confirm that deformation twinning in hcp metals is not the result of a simple shear distortion. Here we present the experimental proof that such an unconventional twin exists; it is millimeter-sized and appears in a bulk magnesium single crystal. It will be also shown that this twin is often co-formed with a new conventional twin on $\{21\bar{3}2\}$ plane that exhibits the lowest shear value ever reported for hcp metals.

A piece of magnesium single crystal was cut with a diamond saw, mechanically polished and then electro-polished. Bands of twins are visible in optical microscopy at the side where the sample was cut. A second cut was performed perpendicularly to the first one, and here again, large bands of twins appeared at the cut side, which shows that the twins were induced by friction during the cutting step. The two cut sections are called A and B in the rest of the paper. Electron BackScatter Diffraction (EBSD) maps were acquired on the area containing the larger twins in the A and B sections. They are shown in Fig. 1 and Extended Data Fig.1, respectively. In order to facilitate the identification of the twins all along this manuscript, some colors were attributed to the different types of twins, independently of the absolute orientation of the sample. The parent single crystal is colored in grey and the conventional extension twins in blue. New twins, colored in green, are formed close to the conventional extension "blue" twins. They are often co-formed with twins colored in yellow, orange and red. The green and yellow/orange/red twins are twins that have never been reported in literature.

Before detailing the crystallographic characteristics of these new twins, it is worth recalling, with the example of the conventional extension blue twins of Fig. 1, how crystallographic information can be read from the experimental EBSD data and their associated pole figures. The extension twins are identified in the EBSD maps by their (86°, **a**) misorientations with the parent crystal, as shown by the rotation of 86° around the **a**-axis marked by the dashed circle in Fig. 2b and c. The atomic displacements during extension twinning are such that the basal {001} plane and prismatic {100} plane are exchanged, as illustrated by the exchange between of the positions of the planes marked by the triangles and squares in the pole figures shown in Fig. 2a and b. The habit plane of the extension twins is the "diagonal" {012} planes located between these two planes. Indeed, the traces of the habit planes agree perfectly with the expected {012} planes, as shown by the fact that spots marked by circles in Fig. 2d are perpendicular to the trace of the extension twins noted HPE1 and HPE2 in Fig. 1. The same results are obtained with the EBSD map acquired on section B, shown in Extended Data Fig.1, with pole figures in Extended Data Fig.2. The {012} habit plane of the extension twins appears in the EBSD map as a plane that is both untilted and undistorted, i.e. fully invariant, in agreement with the simple shear theory, but actually, if the process of lattice distortion is considered in its continuity, the atomic displacements are such that the {012} plane cannot be maintained invariant during the distortion; it is only restored when the distortion is complete[21]. The conventional extension blue twins do not provide a direct footprint of this continuous process, and the {012} interface appears as if the {012} plane had been invariant through the process. The situation will be shown to be different with the "green" twins.

The long millimeter-sized green twins shown in the EBSD map of Fig. 1 are misoriented from the parent crystal by a rotation angle ~58° with a spreading of ± 4° and a rotation axis close to **a** + 2**b**. This misorientation appears in the histogram of Fig. 1b and c. Twins with similar misorientations already appeared in the histograms of some previous studies [25,26], but their crystallographic analysis is very recent[27,28]. Ostapovets *et al.*[27] interpret them as the result of a complete double {012}-{012} twinning in which there is no retained traces of the first {012} twins. A different mechanism was proposed in which the lattice distortion is modelled as a one-step process without the need of a hypothetical intermediate {012} twin[28]. Whatever the model, it is agreed[27,28] that there is a strong link between these (58°, **a**+2**b**) twins and the (64°, **a**+2**b**) {11$\bar{2}$2} twins frequently observed in titanium, and that these twins are not predicted by the general theory of twinning[8,9] or by the dedicated Westlake-Rosembaum model[29,30]. Following our study[28], it will be assumed that the (58°, **a**+2**b**) twins result from a unique distortion that is geometrically represented with the supercell X$_2$YG shown in Fig. 3a. When the parent lattice is rotated by (58°, **a**+2**b**) the supercell becomes close to the initial one, as illustrated in the projection along the axis **OY** = **a**+2**b** of Fig. 3b. The (58°, **a** + 2**b**) prototype configuration is special because the parent vector $\mathbf{OX}_2 = [200]_p$ is parallel to the twin vector $\mathbf{OX}_2{'} = [\bar{1}01]_{gr}$, the parent vector $\mathbf{OY} = [120]_p$ is invariant, and the parent vector $\mathbf{OG} = [\bar{1}01]_p$ is parallel to the twin vector $\mathbf{OG'} = [200]_{gr}$, with the indices "p" and "gr" given in reference to the parent and green twin bases. In addition to the parallelism of the directions, the lengths of the vectors are such that $\|\mathbf{OX}_2\| \approx \|\mathbf{OX'}_2\|$, $\|\mathbf{OY}\| = \|\mathbf{OY'}\|$ and $\|\mathbf{OG}\| \approx \|\mathbf{OG'}\|$. The (p → gr) distortion associated with the transformation $\mathbf{OX}_2 \rightarrow \mathbf{OX'}_2$, $\mathbf{OY} \rightarrow \mathbf{OY'}$, $\mathbf{OG} \rightarrow \mathbf{OG'}$ is given by a upper triangular matrix $\mathbf{F}_{hex}^{p \rightarrow gr}$ given in **Supplementary equations (15)**, with details reported elsewhere[28]. Its diagonal values are $\frac{\sqrt{1+\gamma^2}}{2}$, 1 and $\frac{2}{\sqrt{1+\gamma^2}}$, where $\gamma$ is the c/a packing ratio. The principal strains associated with this distortion are -4.2%, 0, and +4.4% in the case of pure magnesium. The parent/twin misorientation matrix is a rotation of axis **OY** = $[120]_p$ and of angle $ArcTan(\gamma)$. It takes the value 58.39° for magnesium; which is expected from the model and very close to the misorientation observed in the histogram of Fig. 1b and in Fig. 4a and b. The correspondence matrix $\mathbf{C}_{hex}^{gr \to p}$ calculated by considering the vectors $\mathbf{OX}_2, \mathbf{OY}$, and $\mathbf{OG}$, and the vectors $\mathbf{OX}_2', \mathbf{OY}'$, and $\mathbf{OG}'$ expressed in their respective hexagonal bases is given in **Supplementary Equations (17)**. The same correspondence expressed in the reciprocal space $\left(\mathbf{C}_{hex}^{gr \to p}\right)^*$ is the inverse of the transpose of $\mathbf{C}_{hex}^{gr \to p}$; it is in agreement with the fact that the {001} and {112} planes are interchanged by the twin, as illustrated by the exchange between the triangles and squares in the pole figures shown in Fig. 4b and c. The details of the calculations of distortion, orientation and correspondence matrices are given in section 2 of **Supplementary Equation** and are computed in part A of the Mathematica program reported in **Supplementary Data**.

The (58°, **a**+2**b**) twins is only a stretch prototype close to the green twins observed in Fig. 1. From a theoretical point of view, its role is similar to the (90°, **a**) stretch prototype used in the atomistic model of the (86°, **a**) extension twin[21], or to the Bain distortion used as a prototype stretch distortion of the fcc-bcc martensitic transformation in the Phenomenological Theory of Martensite Transformation (PTMC)[31]. The usual way to build a conventional twin from the prototype model is to add a small obliquity correction (few degrees) that compensates the tilt of an undistorted plane in order to make it fully invariant. After obliquity correction, the distortion becomes a simple shear. Calculations prove that there are only two possible planes whose tilt can be compensated by an obliquity correction; they are the planes $(2\bar{1}2)$ and $(\bar{1}26)$ as detailed in separate papers[27, 28]. In both cases, the rotation axis of the obliquity compensation is $\mathbf{OY} = [120]_p$. Geometrically, the $(2\bar{1}2)$ and $(\bar{1}26)$ are the two diagonals of the OX$_2$VG rhombus shown in Fig. 3c and d, respectively; they define two conjugate twinning modes. The shear vector is along the $[101]$ direction in the $(2\bar{1}2)$ twin mode, and along the $[\bar{3}\,0\,1]$ direction in the $(\bar{1}26)$ twin mode, as illustrated by the green arrows in Fig. 3c and d. The shear magnitudes are the same for both modes and close to 0.11 for magnesium[27,28]. The new parent-twin misorientation of the $(2\bar{1}2)$ twin is a (63°, **a**+2**b**) rotation (the corrected obliquity was 4°), and the new parent-twin misorientation of the $(\bar{1}26)$ twin is a (57°, **a**+2**b**) rotation (the corrected obliquity was 1°). These two twins modes are not reported in the list of twins predicted by the classical shear theory[9]; they are however conventional because their lattice distortions are given by simple shear matrices. By considering these theoretical results, one could expect that the habit planes of the green twins, noted HP1 and HP2 in the EBSD maps of Fig. 1, and those of the twins in the EBSD map of Extended Data Fig.1, are two equivalent planes in the family of the $\{11\bar{2}2\}$ planes, or in the family of the $\{11\bar{2}6\}$ planes. Surprisingly, this is not the case. The two HP1 and HP2 green twins have very close orientations (their disorientation angle is lower than 4°) but very distinct habit planes, whereas the models[27,28] predict only one habit plane per family. In the pole figures of Fig. 4, the unique plane of type $\{11\bar{2}6\}$ and the unique plane of type $\{11\bar{2}2\}$ common to both the parent crystal and the green twin are encircled; and, as they are close to the **x**-axis, their trace on the EBSD map should be vertical in the EBSD map of Fig. 1, which is clearly not the case (they are at more than 50° away from the vertical direction). So, what are the habit planes of the green twins? After some attempts we discovered that they are $\{212\}_p$ planes of the parent crystal *and* $\{012\}_{gr}$ planes of the green twin crystals. This is shown in Fig. 5a and b by the fact that (*i*) the circles around the $\{012\}_t$ pole are positioned exactly at the same positions as those of the $\{212\}_p$ poles, and (*ii*) these common poles are perpendicular to the traces of the habit planes HP1 and HP2 of the green twins shown in the EBSD map of Fig. 1. The $\{212\}_p \rightarrow \{012\}_{gr}$ correspondence is expected by the theoretical correspondence matrix $\left(\mathbf{C}_{hex}^{t \rightarrow p}\right)^*$, as shown by **Supplementary Table 2** in **Supplementary Equation**; but the EBSD experimental results bring an additional information: the planes $\{212\}_p$ and $\{012\}_{gr}$ are exactly parallel, which is not the case by considering only the theoretical prototype (58°, **a**+2**b**) stretch distortion given in **Supplementary Equations (15)**. This experimental result is of prime importance because contrarily to all deformation twins ever reported in bulk materials, the habit planes of the green twins are not invariant planes; they cannot result from a simple shear. They are thus called here "unconventional", and noted ~(58°, **a**+2**b**). Additional crystallographic information is required to get a better understanding of these new twins. It was found that that the green twins share with the parent crystal a common direction of type ⟨201⟩, i.e. ⟨22$\bar{4}$3⟩ in the four index Miller-Bravais notation, and that this direction, marked by a blue circle in Fig. 5c, lies in the habit plane whose normal direction is encircled in red in Fig. 5b. The EBSD map acquired on section B and reported in Extended Data Fig.1 with pole figures in Extended Data Fig.3 presents green twins with exactly the same crystallographic characteristics as those of Fig. 1. Some TEM observations were also realized on a green twin; they do not reveal additional internal twinning or important dislocation arrays, as shown by the TEM bright field image in Extended Data Fig.4.

Let us build a crystallographic model of the ~(58°, **a**+2**b**) green twins. The parent/twin misorientation should be close (within few degrees) to that of the (58°, **a**+2**b**) stretch prototype previously discussed. Their habit plane is the $\boldsymbol{g}_0 = (212)_p$ plane that is transformed into the $(012)_{gr}$ plane. The $(212)_p \rightarrow (012)_{gr}$ distortion occurs by the displacements of the atoms located in the upper layer *l* = 1/3 of the plane $(212)_p$ as shown in in Fig. 6. In the $(212)_p$ plane, the direction $[\bar{1}01]_p$ is 4% elongated to be transformed into $[\bar{2}00]_{gr}$, and the angle $\alpha_p = ([\bar{1}01]_p, [0\bar{2}1]_p)$ decreases by 3°. The details of the calculations are in section 2 of **Supplementary Equations**. Now, the distortion should be obliquity-corrected such that the plane $\boldsymbol{g}_0$ and the direction $\boldsymbol{u}_0 = [0\bar{2}1]$ are simultaneously maintained unrotated. This direction $[0\bar{2}1]$ was chosen among the equivalent ⟨22$\bar{4}$3⟩ directions because it lies in the $(212)_p$ plane. From these conditions and the initial model of the (58°, **a**+2**b**) prototype, it is now possible to build a crystallographic model of the green twins. The calculations are detailed in section 2 of **Supplementary Equations** and are computed in part B of the Mathematica program reported in **Supplementary Data**. They prove that the obliquity angle is 3.3°. The nine components of the obliquity-corrected distortion matrix, as function of the packing ratio, are given in **Supplementary Equation (23)**. The new parent/twin misorientation of the obliquity-corrected twin is a rotation given in **Supplementary Equation (25)**; the rotation angle is 60.7° and the rotation axis is less than 2° away from the **a**+2**b** axis. The difference of the orientation between the twins formed directly by the prototype model and those formed with the obliquity-corrected version is low (58.4° / 60.7°) and lies in the spreading of the misorientation histogram shown in Fig. 1b. This spreading exists for all observed green twins. It was noticed that the isolated green twins have a greater tendency to exhibit a 58° misorientation with the parent crystal, such as the one marked by green arrow in Fig. 1a, whereas the green twins that are co-formed with yellow twins tend to exhibit misorientations more centered in the range 60-61°; it is the case in the area marked by dashed green rectangle in Fig. 1a, as shown by the local misorientation histogram given in Fig. 1e. Gradient of orientations between 58° and 62° are rainbow-colored in Fig. 1d. The yellow twins seem to stabilize the unconventional green twins such that the condition $\boldsymbol{g}_0 = (212)_p \mathbin{/\mkern-5mu/} (012)_{gr}$ is fulfilled. The following part is devoted to the crystallographic properties of the yellow twins and their role in the stabilization of the green twins.

The yellow twins have a misorientation with the green twins with which they are co-formed close to (86°, **a**). This shows that the yellow twins are linked to the green twins by a sort of extension twinning. However, the yellow twins, as the green twins, also result from a deformation twinning mechanism of the parent crystal. The misorientation of the yellow twins with the parent crystal is found to be close to a rotation of 48° around an axis of type <241>, as illustrated in Fig. 1b and c for the EBSD map of section A, and in Extended Data Fig.1b and c for the EBSD of the section B. The habit planes with the parent crystal are the same as those of the green twin; they are $\{212\}_p$ planes. Consequently, contrarily to the green twins for which the habit plane is a distorted plane $\{212\}_p \rightarrow \{012\}_{gr}$, the habit plane of the yellow twin is a restored plane $\{212\}_p \rightarrow \{212\}_{ye}$, as shown in the pole figures of Fig. 5b. Indeed, in this figure, the circle noted "HP1" is around the {212} plane that is common to both the parent and the yellow twins co-formed with the green twin lying along HP1 in Fig. 1; and the circle noted "HP2" is around the {212} plane that is common to both the parent and the red twin co-formed with the green twin lying along HP2 in Fig. 1. These features are also observed with the three habit planes identified in the EBSD map of section B, as shown in the {212} pole figure of Extended Data Fig.3 of the yellow, orange and red twins co-formed with the green twins shown in Extended Data Fig.2. Thus, the yellow twins are conventional because their habit plane is a crystallographic plane that is common to both parent and twin crystals. However, to the best of our knowledge, no twin with a shear plane of type $\{21\bar{3}2\}$ has ever been reported by the classical models of deformation twinning[9]. It is thus important to get additional crystallographic information on this twin and its link with the green twin. It was noticed that the axis $\boldsymbol{u}_0 = [0\bar{2}1]$ that was left invariant during the $(p \rightarrow gr)$ twinning is also left invariant by the $(gr \rightarrow ye)$ twinning, i.e. this axis is common to the three crystals: the parent crystal, the green twin and the yellow twin. This is shown by the encircled directions in the pole figure Fig. 5c and in the pole figure of Extended Data Fig.3c.

Let us build a crystallographic model of the yellow twins. One could have imagined building such a model by considering that the yellow twins are extension twins of the green grains. However, it is mathematically impossible to build a conventional twin by composing the distortion matrix of the ~(58°, **a**+2**b**) unconventional green twin with that of a conventional (86°, **a**) extension twin because the terms in the former are irrational and those of the latter are rational, which means that their composition cannot give a rational matrix. This apparent issue is solved by considering that the yellow twins are not in a conventional extension twin relationship with the green grains, but with a relation that is derived from it by a small obliquity correction. The green twins induce a planar distortion $(212)_p \rightarrow (012)_{gr}$, and the obliquity-corrected extension twin should be such that it induces the reverse planar distortion $(012)_{gr} \rightarrow (212)_{ye}$. The idea is therefore to maintain untilted the plane $(012)_{gr}$ and invariant the direction $\boldsymbol{u}_0 = [0\bar{2}1]$ during $(gr \rightarrow ye)$ twinning. The calculations are detailed in section 3 of **Supplementary Equations**. They show that an obliquity of 1.1° around the common axis $\boldsymbol{u}_0$ is sufficient to make the conventional extension twins compatible with the experimental results. This obliquity is so small that it is not possible to distinguish whether the green/yellow relation is a conventional (86°, **a**) twinning or its derived version. The new distortion matrix and the misorientation matrix associated with the $(gr \rightarrow ye)$ relation are given in **equation (40) and (42)** of **Supplementary Equations**.

Now, it is possible to compose the correspondence, distortion and orientation matrices of the $(p \rightarrow gr)$ twin with those of the $(gr \rightarrow ye)$ twin. The calculations are detailed in section 4 of **Supplementary Equations**. The calculated misorientation between the yellow twin and the parent crystal is a rotation of 48.7° around the $[\bar{2}21]_p$ axis, which is in perfect agreement with the 48° misorientation experimentally determined in the histograms of Fig. 1b and c, and those of Extended Data Fig.1b and c. The distortion matrix is a shear matrix, as expected for a conventional twin. The shear vector is $[\bar{5}\bar{4}7]_p$, i.e. of type $\langle 12\bar{3}7 \rangle_p$. The shear value is $s = \sqrt{\frac{7}{48} \frac{|3-\gamma^2|}{\gamma}}$, which is equal to 0.078 for ideal hard-sphere packing and 0.084 for magnesium. This twinning mode was not predicted by the classical theory of deformation twinning; the shear magnitude is, to our best knowledge, the lowest value ever reported for deformation twinning of metals.

In summary, the EBSD study on a saw-cut magnesium single crystal put in evidence unconventional millimeter-size twins localized close to conventional extension twins. The parent/twin misorientation is ~(58°, **a**+2**b**). This twin is unconventional because its habit plane is not invariant; it is a {212} plane untilted but distorted and transformed into a {012} plane. This twin is often co-formed with another twin and linked to it by an unconventional type of extension twinning. This co-formed twin is a conventional (shear) twin of the parent crystal, but the twin mode is new; the parent/twin disorientation is (48°,⟨$\bar{2}$21⟩); and the calculations show that the associated distortion matrix is a {212} ⟨$\bar{5}\bar{4}$7⟩ shear with a magnitude of only 0.084. Some researchers recently proposed a "pure shuffle" model to interpret their observations of (90°, **a**) extension twins in magnesium nano-pillars, but they assumed that their discovery was limited to this special twin and was not a "*failure of the classical theory*". The present evidence of macroscopic unconventional twins with {212} → {012} interface calls for reconsidering the theory of deformation twinning because the initial paradigm of simple shear is not consistent with the present observations. An approach based on hard-spheres had been followed for the last years and applied to martensitic transformations and to extension and compression twinning in magnesium[21-24]; it proposes to shift the shear paradigm while preserving the essential displacive features of these transformations[32]. Once generalized and formalized, it could constitute one of the possible alternatives to the shear-based theories.

## Methods

The magnesium single crystal was bought at Goodfellow Inc. It is the same sample as the one used in the theoretical study of extension twinning[21]. Two perpendicular cross-sections were cut and called A and B in the paper. The extension twins and the new twins studied are induced by the disk cutting with the abrasive disk saw. The two sections were mechanically polished with abrasive papers and clothes with diamond particles down to 1 μm, and then electropolished at 12V with an electrolyte made of 85% ethanol, 5% $HNO_3$ and 10% HCl just taken out of the fridge (10°C). The EBSD map was acquired on a field emission gun (FEG) XLF30 scanning electron microscope (FEI) equipped with an Aztec system (Oxford Instruments). The EBSD maps were treated with the Channel5 software (Oxford Instruments). The blue, green and yellow/red colors were attributed to the twins by using the function "Texture Components" with the average Euler angles of each twin and a range angle of 10°.

All the indices of planes and vectors are noted here in the crystallographic hexagonal basis, even if the calculations detailed in **Supplementary Equations** sometimes use intermediate orthonormal bases. The ratio of lattice parameter is $\gamma = \sqrt{8/3}$ for ideal hard-sphere packing and $\gamma = 1.625$ for magnesium. All the calculations were performed in symbolic form with Mathematica; the computer program is given in **Supplementary Data**. The calculations are detailed in **Supplementary Equations**; which includes at its end a table (**Supplementary Table 2**) that summarizes the main equations used in the paper. The vectors are noted by bold lowercase letters and the matrices by bold capital letters. The three-index Miller notation in the hexagonal system is used for the calculations and preferentially chosen to write the results. The four index Miller-Bravais notation is sometimes written to help the reader to identify the directions or planes that are equivalent by hexagonal symmetries. Conversion rules[33] between the three-index and four-index notations for planes and directions are $(h,k,l) = (h,k,\overline{h+k},l)$ and $[u,v,w] = [\frac{2u-v}{3}, \frac{2v-u}{3}, \frac{\overline{u+v}}{3}, w]$, respectively. For example, **a** = [100]$_{hex}$, **b** = [010]$_{hex}$ belong to the set of equivalent directions $\frac{1}{3}\langle 11\overline{2}0 \rangle$. The distortion matrix, the correspondence matrix, and the coordinate transformation matrix are defined in **Supplementary Equations** §1. The crystallographic calculations of the (58°, **a**+2**b**) twin with its obliquity-corrected version are developed in **Supplementary Equations** §2. The crystallographic calculations of the (86°, **a**) extension twin with its unconventional version are given in **Supplementary Equations** §3. The composition of these two unconventional twins leading to the new conventional twin is detailed in **Supplementary Equations** §4. The calculations can be checked by consulting the Mathematica program in **Supplementary Data**.

## References

1. Thomson, W. & Tait, P.G. *Treatise on Natural Philosophy*, Cambridge, vol. I (1) 170-171, pp. 105-106 (1867).

2. Mügge, O. Ueber homogene Deformationen (einfache Schiebungen) an den triklinen Doppelsalzen BaCdCl4.4aq., *Neues Jahrbuch für Mineralogie, Geologie und Palaeontologie*, Beilage-Band **6**, 274–304 (1889).

3. Hardouin Duparc, O.B.M. A review of some elements for the history of mechanical twinning centred on its German origins until Otto Mügge's K1 and K2 invariant plane notation. *J. Mater. Sci.* DOI: 10.1007/s10853-016-0513-4 (2017).

4. Cahn, R.W. Twinned crystals. *Advanced in Physics* **3**, 363-444 (1954).

5. Kihô, H. The Crystallographic Aspect of the Mechanical Twinning in Metals. *J. Phys. Soc. Japan* **9**, 739-747 (1954).

6. Jaswon, M.A. & Dove, D.B. The Crystallography of Deformation Twinning. *Acta Cryst.* **13**, 232-240 (1960).

7. Bilby, B.A. & Crocker, A.G. The Theory of the Crystallography of Deformation Twinning. *Proc. R. Soc. Lond. A*. **288**, 240-255 (1965).

8   Bevis, M. & Crocker, A.G. Twinning Shears in Lattices. *Proc. Roy. Soc. Lond. A*. **304**, 123-134 (1968).

9   Crocker, A.G. & Bevis, M. *The Science Technology and Application of Titanium*, ed. R. Jaffee and N. Promisel, Pergamon Press, Oxford, pp. 453-458 (1970).

10  Frenkel, J. Zur theorie der elastizit. atsgrenze und der festigkeit kristallinischer kцrper. *Z. Phys.* **37**, 572-609 (1926).

11  Cottrell, A.H. & Bilby, B.A. A mechanism for the growth of deformation twins in crystals. *Phil. Mag*. **42** 573-581 (1951).

12  Sleeswyk, A.W. Perfect dislocation pole models for twinning in f.c.c and b.c.c lattices. *Phil. Mag.* **29**, 407-421 (1974).

13  Venables, J.A. A dislocation pole for twinning, *Phil. Mag.* **30**, 1165-1169 (1974).

14  Hirth, J.P. & Pond, R.C. Steps, Dislocations and disconnections as interface defects relating to structure and phase transformations. *Acta Mater.* **44**, 4749–4763 (1996).

15  Pond, R.C., Hirth, J.P., Serra, A. & Bacon, D.J. Atomic displacements accompanying deformation twinning: shear and shuffles. *Mater. Res. Lett.* DOI: 10.1080/21663831.2016.1165298 (2016)

16  Liu, B.-Y. *et al.* Twinning-like lattice reorientation without a crystallographic plane. *Nat. Com*. **5**, 3297 (2014).

17  Li, B. & Ma, E. Atomic Shuffling Dominated Mechanism for Deformation Twinning in Magnesium. *Phys. Rev. Lett.* **103**, 035503 (2009).

18  Wang, J. *et al.* Pure-Shuffle Nucleation of Deformation Twins in Hexagonal Close-Packed Metals. *Mater. Res. Lett.* **1**, 126-132 (2013).

19  Li, B. & Zhang, X.Y. Twinning with zero shear. *Scripta Mater.* **125**, 73-79 (2016).

20  Tu, J. & Zhang, S. On the {10$\bar{1}$2} twinning growth mechanism in hexagonal close-packed metals. *Mater. Design* **96**, 143-149 (2016).

21  Cayron C. Hard-sphere displacive model of extension twinning in magnesium. *Mater. Design* **119**, 361-375 (2017).

22  Cayron, C. Hard-sphere displacive model of deformation twinning in hexagonal close-packed metals. Revisiting the case of the (56°, a) contraction twins in magnesium. *Acta Cryst. A* **73**, https://doi.org/10.1107/S2053273317005459 (2017).

23  Hales, T.C. A proof of the Kepler conjecture, *Annals of Mathematics. Second Series*, **162**, 1065-1185 (2005).

24  Cayron, C. Angular distortive matrices of phase transitions in the fcc-bcc-hcp system. *Acta Mater.* **111**, 417-441 (2016).

25  Nave, M.N. & Barnett, M. R. Microstructure and textures of pure magnesium deformed in plane-strain compression. *Scripta Mater.* **51**, 81-885 (2004).

26  Lentz, M. *et al.* In-situ, ex-situ and (HR-)TEM analyses of primary, secondary and tertiary twin development in an Mg-4wt%Li alloy. *Mater. Sci. Engng A* **610**, 54-64 (2014).

27  Ostapovets, A. *et al.* On the relationship between $\{11\bar{2}2\}$ and $\{11\bar{2}6\}$ conjugate twins and double extension twins in rolled pure Mg. *Phil. Mag.* **97**, 1088-1101 (2017).

28  Cayron, C. The $(11\bar{2}2)$ and $(\bar{1}2\bar{1}6)$ twinning modes modelled by obliquity correction of a (58°, **a**+2**b**) prototype stretch twin. https://arxiv.org/ftp/arxiv/papers/1706/1706.08338.pdf

29  Westlake, D.G. Twinning in zirconium. *Acta Metall.* **9**, 327-331 (1961).

30  Rosenbaum, H.S. *Nonbasal Slip in h.c.p. Metals and its Relation to Mechanical Twinning, in Deformation Twinning*, ed. By R.E. Reed-Hill, J.P. Hirth, H.C. Rogers, New York: Gordon & Breach; pp 43-76 (1964).

31  Bhadeshia, H.K.D.H. *Worked examples in the geometry of crystals.* $2^d$ ed. Brookfield, The Institute of Metals (1987).

32  Cayron, C. Over the shear paradigm, https://arxiv.org/ftp/arxiv/papers/1706/1706.07750.pdf

33  Partridge, P.G. The crystallography and deformation modes of hexagonal close-packed metals. *J. Metall. Reviews* **12**, 169-194 (1967).

## Acknowledgments

We would like to show our gratitude to PX group for the laboratory funding and for our scientific and technical exchanges.

## Contributions

CC is at the origin of the work; he prepared the SEM and TEM samples, acquired the EBSD maps and the TEM images, interpreted the results, made the crystallographic model, and wrote the manuscript. RL significantly contributed to improve the readability of the manuscript. CC and RL discussed the results and commented on the models.

## Author information

Correspondence and requests for materials should be addressed to C.C. (cyril.cayron@epfl.ch)

# Figures

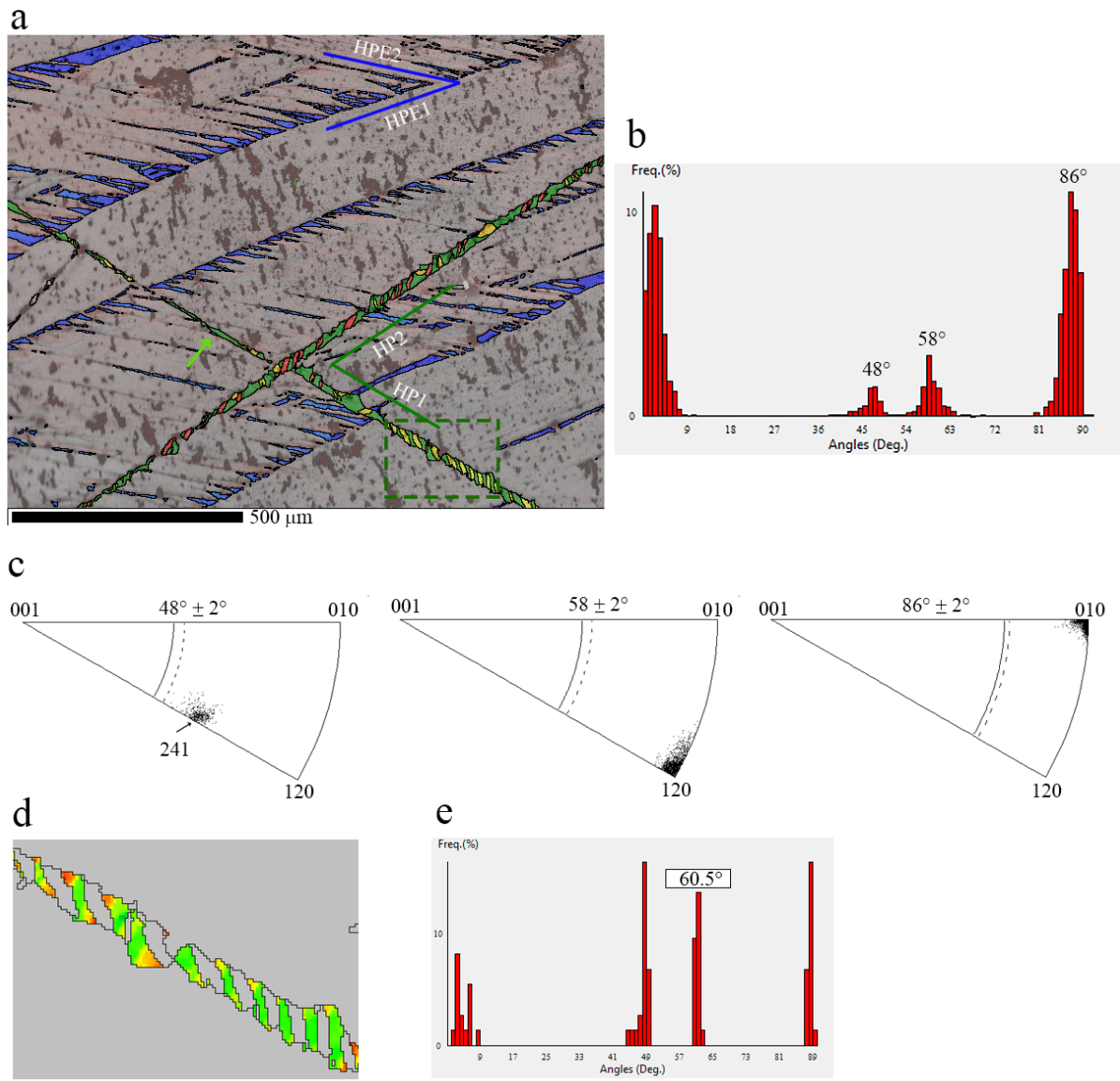

Fig. 1. EBSD map on section A of the magnesium single crystal. (a) Map with colors chosen according to the twin type: the parent crystal is in grey, the (86°, a) extension twins are in blue, the ~(58°, a+2b) unconventional twins are in green, and the (48°, <241>) twins are in yellow and red. (b) Disorientation histogram, with in (c) the rotation axes corresponding to the three peaks of the histogram plotted in the fundamental sector of the hexagonal lattice. The green twins can be found isolated, as marked by the green arrow, or co-formed with the yellow-red twins, as marked by the green rectangle. (d) Enlargement of a co-formation of green and yellow twins, with rainbow colors chosen to amplify the internal orientation gradients in the range (58°-64°). (e) Disorientation histogram of the zone (d) showing an average disorientation between the green twin and the parent crystal at 60.5°, and not 58° as in the rest of the EBSD map.

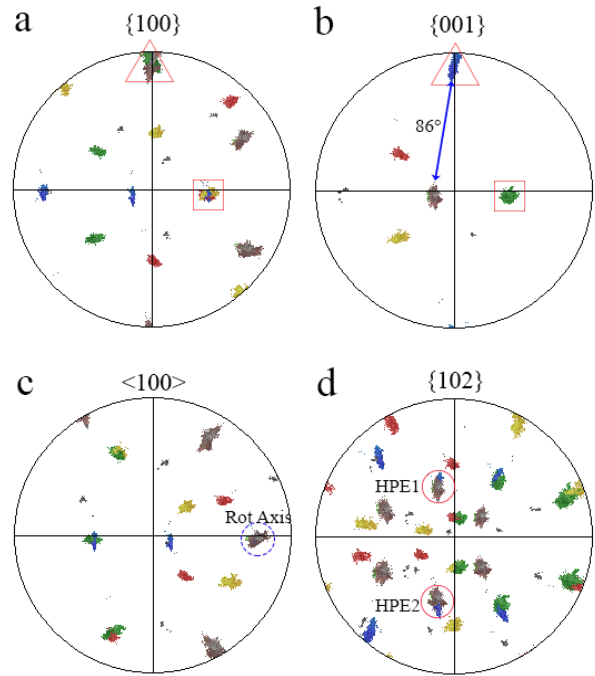

*Fig. 2. Pole figures of the map EBSD shown in Fig. 1 indicating some important crystallographic planes and directions related to the (86°, **a**) extension twins. The correspondence between the basal and prismatic planes is shown by the similar positions of the red rectangles and red squares in (a) and (b). The rotation axis between the parent and the twin is the **a**-axis marked by the dashed blue circle in (c); the rotation angle is 86°, as shown by the blue line between the two c-axes in (b). The habit planes are the {102} planes marked by the red circles and noted HPE1 and HPE2 in (d). The line between the center of the pole figure and these poles are perpendicular to the traces of the interfaces noted HPE1 and HPE2 in Fig. 1a.*

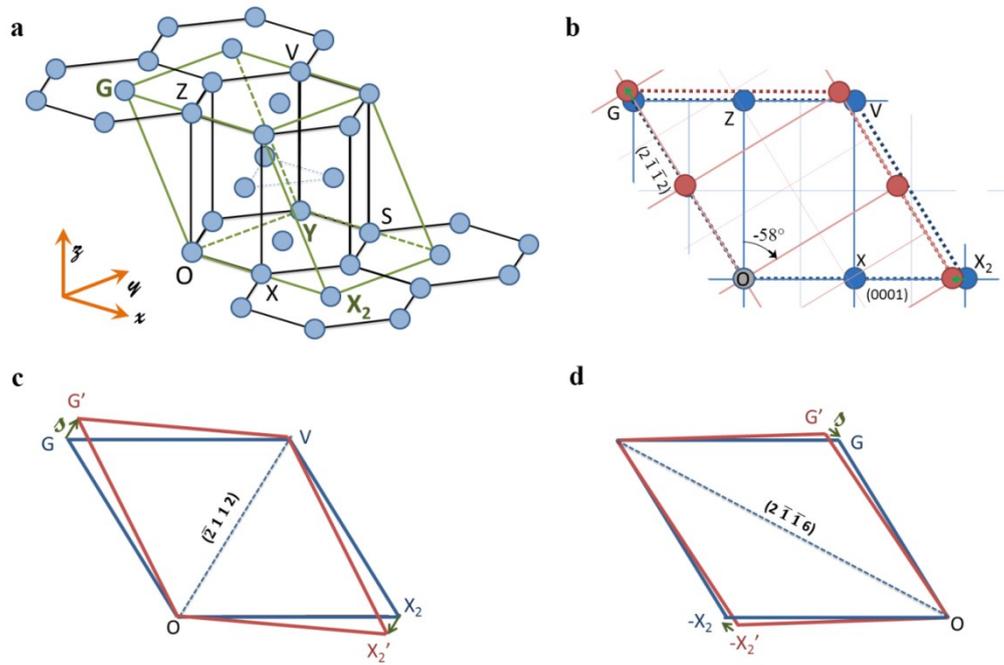

*Fig. 3. Schematic representation of the prototype (58°, **a+2b**) stretch twin. (a) Supercell OX$_2$YG chosen for the calculations. (b) Projection along the **OY** = **a+2b** axis showing that after a (58°, **a+2b**) rotation, the two supercells nearly overlap; the supercell of the parent hcp cell (in blue) can be transformed into the supercell of the twin (in red) by a stretch distortion represented by the green arrows. This twinning mode establishes a correspondence between the basal plane and the $(2\bar{1}2)$ plane. Obliquity corrections are possible to build conventional (shear) twins from the prototype twin; they consists in maintaining untilted an undistorted plane; the two possible planes are: (c) the $(\bar{2}12)$ plane, and (d) the $(2\bar{1}6)$ plane.*

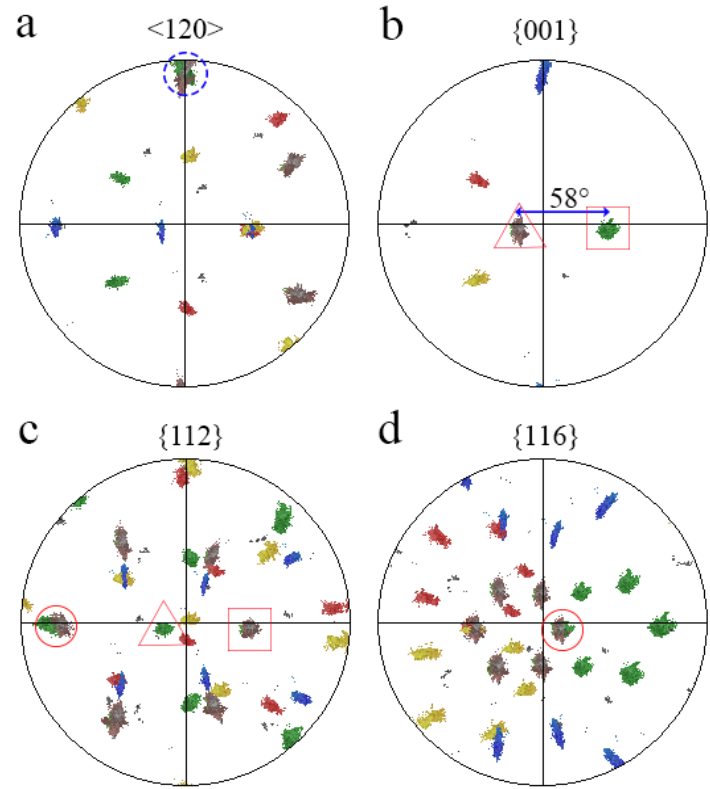

Fig. 4. Pole figures of the EBSD map shown in Fig. 1 indicating some important crystallographic planes and directions related to the unconventional ~(58°, **a**+2**b**) green twin. The correspondence between the basal plane and the {112} planes is shown by the similar positions of the red rectangles and red squares in (b) and (c). The rotation axis between the parent and the twin is the axis <120> marked by the dashed blue circle in (a); the rotation angle is 58°, as shown by the blue line between the two **c**-axes in (b). The two undistorted planes common to the twin and the parent crystal are the {112} and {116} planes marked by the red circles in (c) and (d). None of these two planes agrees with the traces of the habit planes of the green twin noted HP1 and HP2 in Fig. 1.

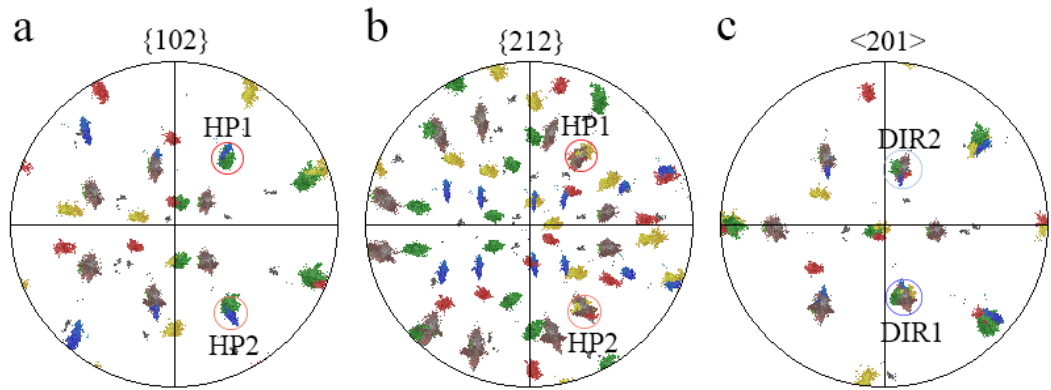

Fig. 5. Pole figures of the EBSD map revealing the unconventional character of the ~(58°, **a**+2**b**) green twin. The trace of the green twins noted HP1 and HP2 in Fig. 1 are perpendicular to the planes {212}$_p$ of the parent crystal marked by the red circles around the grey spots in (b), and these planes are parallel to some of the {102}$_{gr}$ planes of the green twin marked by the red circles around the green spots in (a). It can also be noticed that these {102} planes are common planes of the green and yellow twins. The <201> pole figure in (c) shows that the <201> directions noted DIR1 and DIR2 belonging to the {212} plane HP1 and HP2, respectively, are common to the parent, green and yellow crystals.

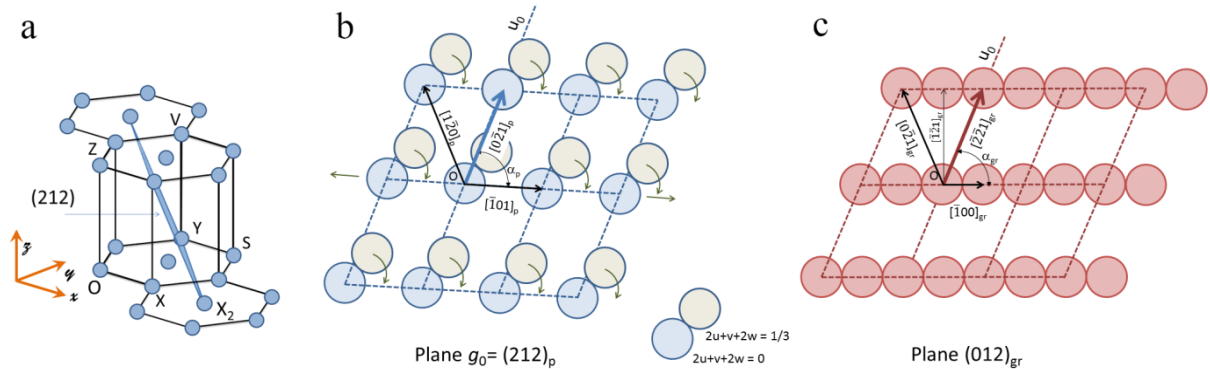

Fig. 6. Atomic model of the (212)$_p$→(012)$_{gr}$ transformation. (a) (212)$_p$ plane of the parent crystal drawn in the reference frame. (b) The same plane viewed edge-on; it is constituted of two parallel layers of atoms, one with atoms of coordinates [u, v, w] positioned at the level l = 2u+v+2w = 0 (in blue), and the other one with atoms at the level l = 2u+v+2w = 1/3 (in light grey). The displacements of the atoms of the layer l = 1/3 down to the lower layer l = 0 are shown by the green curves arrows. (c) Plane (012)$_{gr}$ of the green twin constituted of only one layer l = v+2w = 0 (in red), obtained after the atomic displacements and lattice distortion.

# Extended Data

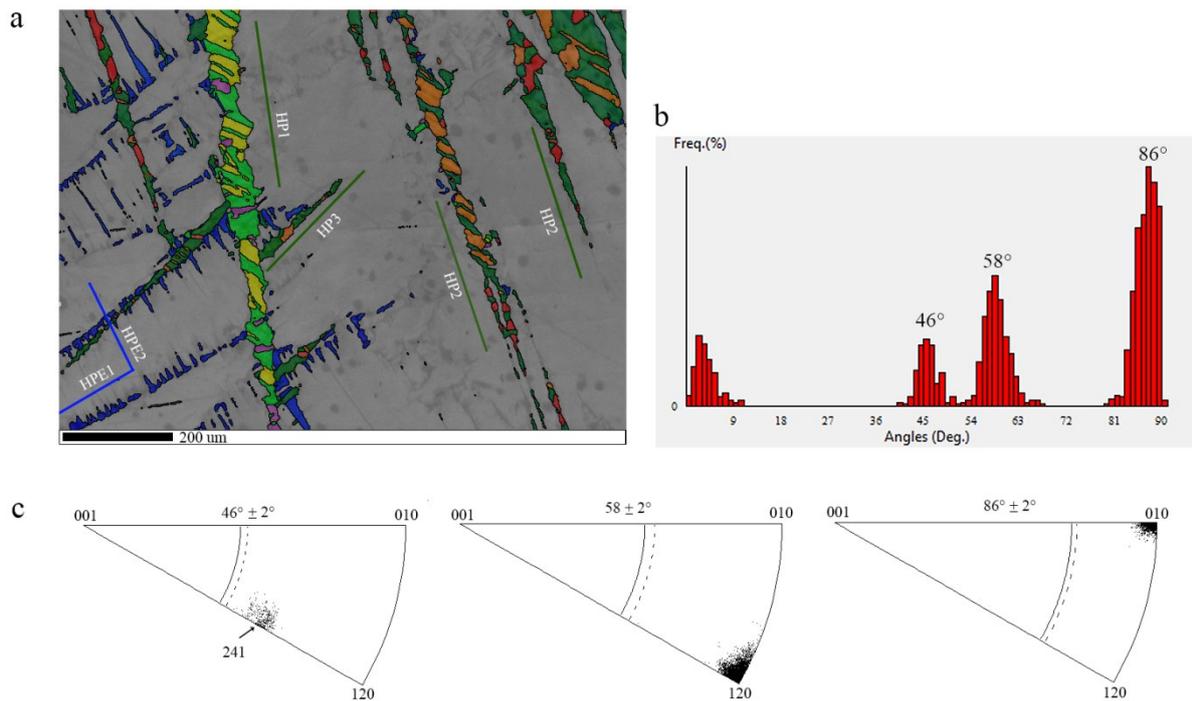

*Extended Data Fig.1. EBSD map of the magnesium single crystal cut on section B. (a) Map with colors chosen according to those used for the different twin types in Fig. 1a. (b) Disorientation histogram, with in (c) the rotation axes corresponding to the three peaks of the histogram plotted in the fundamental sector of the hexagonal lattice.*

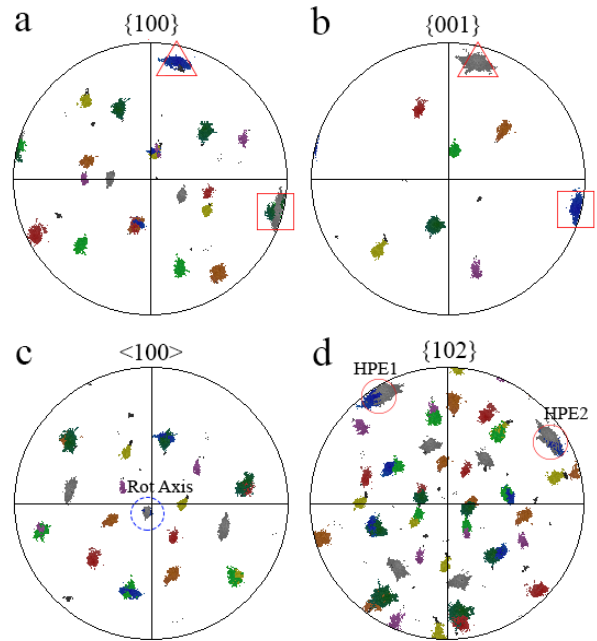

*Extended Data Fig.2. Pole figures of the map EBSD shown in Extended Data Fig.1 indicating some important crystallographic planes and directions related to the (86°, **a**) extension twin. The planes and directions marked by the rectangles, triangles and circles are interpreted exactly as in Fig. 2.*

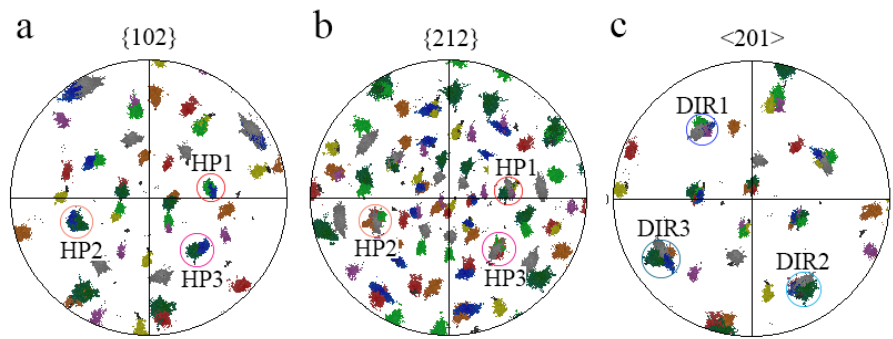

*Extended Data Fig.3. Pole figures of the map EBSD shown in Extended Data Fig.1 confirming the unconventional character of the ~(58°, **a**+2**b**) green twins already shown in Fig. 3. Two variants of ~(58°, **a**+2**b**) are visible in this map (light and dark green): the light green twins exhibit two different habit planes, and the dark green twin only has one habit plane. These three habit planes, noted HP1, HP2 and HP3, are perfectly indexed as $\{212\}_p // \{012\}_{gr}$ planes. It was also checked that all these twins share a common ⟨201⟩ direction, noted DIR1, DIR2 and DIR3, that belongs to the {212} plane HP1, HP2 and HP3, respectively.*

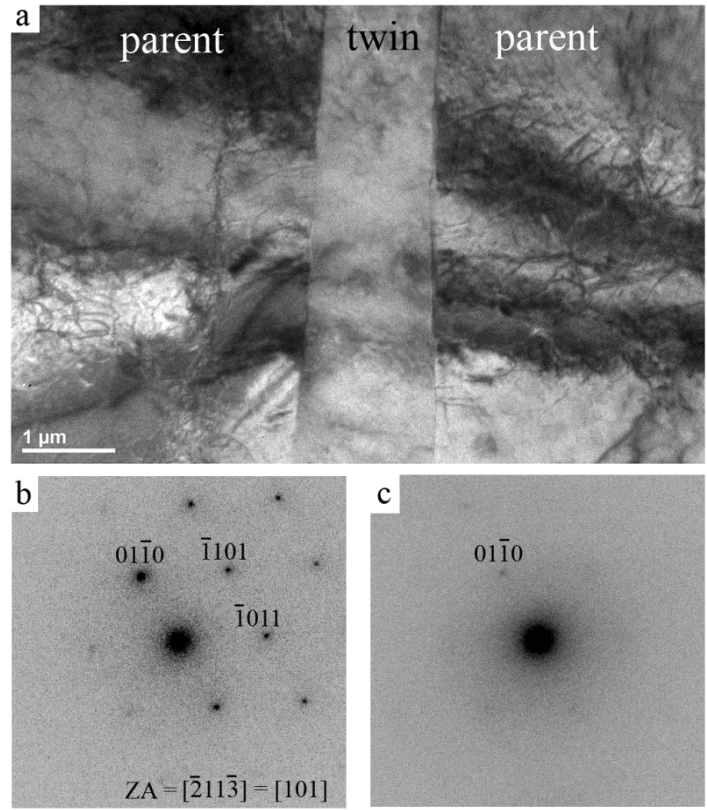

*Extended Data Fig.4.    TEM image of an unconventional ~(58°, **a+2b**) twin. (a) Bright field image, (b) selected area diffraction pattern along the [101]$_{hex}$ zone axis (ZA) of the twin, (c) selected area diffraction pattern in the surrounding parent crystal with the same sample orientation as that of (b). No secondary twins or special dislocation pile-ups were detected in the twin.*

# Supplementary Equations

## 1. Preliminary

### 1.1. Definition of the distortion, orientation and correspondence matrices

Deformation twinning is a lattice transformation under stress or strain from a parent crystal (p) to the twinned crystal (t); this distortion restores the lattice in a new orientation. Mathematically, it can be defined by a *distortion matrix* $\mathbf{D}^{p \to t}$. Any direction $\boldsymbol{u}$ is transformed after distortion into a new direction $\boldsymbol{u}' = \mathbf{D}^{p \to t} \boldsymbol{u}$. A plane $\boldsymbol{g}$, considered as a vector of the reciprocal space, is transformed after distortion into a new plane $\boldsymbol{g}' = \left(\mathbf{D}^{p \to t}\right)^* \boldsymbol{g}$ with $\left(\mathbf{D}^{p \to t}\right)^* = \left(\mathbf{D}^{p \to t}\right)^{-T}$ where the symbol –T means the inverse of the transpose.

It is often necessary for the calculation to switch from the crystallographic basis to an orthonormal basis linked to this basis. In the case of an hexagonal phase, we call $\mathbf{B}_{hex} = (\boldsymbol{a}, \boldsymbol{b}, \boldsymbol{c})$ the usual hexagonal basis, and $\mathbf{B}_{ortho} = (\boldsymbol{x}, \boldsymbol{y}, \boldsymbol{z})$ the orthonormal basis linked to $\mathbf{B}_{hex}$ by the coordinate transformation matrix $\mathbf{H}_{hex}$:

$$\mathbf{H}_{hex} = [\mathbf{B}_{ortho} \to \mathbf{B}_{hex}] = \begin{pmatrix} 1 & -1/2 & 0 \\ 0 & \sqrt{3}/2 & 0 \\ 0 & 0 & \gamma \end{pmatrix} \tag{1}$$

where $\gamma$ is the c/a packing ratio of the hexagonal phase. The matrix $\mathbf{H}_{hex}$ is commonly called structure tensor in crystallography. It can be used to express the directions into the orthonormal basis $\mathbf{B}_{ortho}$. For planes, it is $\mathbf{H}^*_{hex}$ that should be used. We note O, the "zero" position that will be left invariant by the distortion and we note X, Y and Z the atomic positions defined by the vectors **OX** = $\boldsymbol{a}$ = [100]$_{hex}$, **OY** = $\boldsymbol{a}$ + 2$\boldsymbol{b}$ = [120]$_{hex}$ and **OZ** = $\boldsymbol{c}$ = [001]$_{hex}$. It can be checked with the matrix $\mathbf{H}_{hex}$ that **OX** = [100]$_{ortho}$, **OY** = [0 $\sqrt{3}$ 0]$_{ortho}$ and **OZ** = [0 0 $\gamma$]$_{ortho}$.

The vectors of the initial parent basis are transformed by the distortion into new vectors: $\mathbf{a}_p \to \mathbf{a}'_p$, $\mathbf{b}_p \to \mathbf{b}'_p$ and $\mathbf{c}_p \to \mathbf{c}'_p$. The distortion matrix expressed in the hexagonal basis $\mathbf{D}^{p \to t}_{hex}$ is the matrix formed by the images $\mathbf{a}'_p$, $\mathbf{b}'_p$ and $\mathbf{c}'_p$ expressed in $\mathbf{B}_{hex}$, i.e. $\mathbf{D}^{p \to t}_{hex} = \left[\mathbf{B}^p_{hex} \to \mathbf{B}'^p_{hex}\right] = \mathbf{B}'^p_{hex}$ with $\mathbf{B}^p_{hex} = (\mathbf{a}_p, \mathbf{b}_p, \mathbf{c}_p)$ and $\mathbf{B}'^p_{hex} = (\mathbf{a}'_p, \mathbf{b}'_p, \mathbf{c}'_p)$. In simple words, the distortion matrix is expressed by writing in column the coordinates of $\mathbf{a}'_p$, $\mathbf{b}'_p$ and $\mathbf{c}'_p$ in the basis $\mathbf{B}^p_{hex}$. The crystallographic studies on displacive phase transformations and mechanical twinning often consist in finding the distortion matrices close to the identity matrix in order to minimize the atomic displacements.

If the distortion matrix is known in the basis $\mathbf{B}_{ortho}$, and noted $\mathbf{D}^{p \to t}_{ortho}$, a formula of coordinate transformation can be used to express it in the basis $\mathbf{B}_{hex}$ ; it is:

$$\mathbf{D}^{p \to t}_{hex} = \mathbf{H}^{-1}_{hex} \mathbf{D}^{p \to t}_{ortho} \mathbf{H}_{hex} \tag{2}$$

with $\mathbf{H}_{hex}$ given by equation (1). Inversely, if the distortion matrix is found in $\mathbf{B}_{hex}$ and it can be written in $\mathbf{B}_{ortho}$ by the inverse formula:

$$\mathbf{D}_{ortho}^{p \to t} = \mathbf{H}_{hex} \mathbf{D}_{hex}^{p \to t} \mathbf{H}_{hex}^{-1} \qquad (3)$$

The *misorientation matrix* is defined by the coordinate transformation matrix $\mathbf{T}_{hex}^{p \to t}$. This matrix allows the change of the coordinates of a fixed vector between the parent and twin bases. It is given by the vectors forming the basis of the twin $\mathbf{B}_{hex}^t = (\mathbf{a}_t, \mathbf{b}_t, \mathbf{c}_t)$ expressed in the parent hexagonal basis, i.e. $\mathbf{T}_{hex}^{p \to t} = [\mathbf{B}_{hex}^p \to \mathbf{B}_{hex}^t]$. Its reverse is just $\mathbf{T}_{hex}^{t \to p} = [\mathbf{B}_{hex}^t \to \mathbf{B}_{hex}^p]$.

The orientation of the twinned crystal is defined by the matrix $\mathbf{T}_{hex}^{p \to t}$, but other equivalent matrices could be chosen. The equivalent matrices are obtained by multiplying $\mathbf{T}_{hex}^{p \to t}$ by the matrices $g_i$ of internal symmetries of the hexagonal phase, i.e. the matrices forming the point group of the hcp phase $\mathbb{G}^{hcp}$.

$$\{\mathbf{T}_{hex}^{p \to t}\} = \{\mathbf{T}_{hex}^{p \to t} g_i, \; g_i \in \mathbb{G}^{hcp}\} \qquad (4)$$

The matrix $\mathbf{T}_{hex}^{p \to t}$ is a coordinate transformation matrix between two hexagonal bases; it is thus a rotation matrix. The rotation angle of a matrix $\mathbf{T}_{hex}^{p \to t}$ is given by its trace and the rotation axis is the eigenvector associated with the unit eigenvalue. However, one must keep in mind that $\mathbf{T}_{hex}^{p \to t}$ is expressed in a non-orthonormal basis, which implies that some usual equations related to rotations do not hold. For example, the inverse of a rotation matrix equals its transposes only in orthonormal basis. Using $\mathbf{T}_{ortho}^{p \to t} = \mathbf{H}_{hex} \mathbf{T}_{hex}^{p \to t} (\mathbf{H}_{hex})^{-1}$ in the calculations allow avoiding possible errors.

In the set of equivalent matrices $\{\mathbf{T}_{hex}^{p \to t}\}$, it is custom to choose the rotation with the lowest angle, called "disorientation". This choice has practical applications, but it remains arbitrary.

The *correspondence matrix* $\mathbf{C}_{hex}^{t \to p}$ gives the distortion images expressed in the twin basis of the parent basis vectors, i.e. $\mathbf{a}'_p$, $\mathbf{b}'_p$ and $\mathbf{c}'_p$. These images are obtained from the misorientation matrix and the distortion matrix: $(\mathbf{a}'_p, \mathbf{b}'_p, \mathbf{c}'_p)_{/\mathbf{B}_{hex}^t} = \mathbf{T}_{hex}^{t \to p} (\mathbf{a}'_p, \mathbf{b}'_p, \mathbf{c}'_p)_{/\mathbf{B}_{hex}^p} = \mathbf{T}_{hex}^{t \to p} \mathbf{B}_{hex}^{\prime p} = \mathbf{T}_{hex}^{t \to p} \mathbf{D}_{hex}^{p \to t}$. The correspondence matrix is thus:

$$\mathbf{C}_{hex}^{t \to p} = \mathbf{T}_{hex}^{t \to p} \mathbf{D}_{hex}^{p \to t} \qquad (5)$$

The correspondence matrix is used to calculate in the twin basis the coordinates of the image by the distortion of a vector written in the parent basis, i.e.

$$\mathbf{x}'_{/\mathbf{B}_{hex}^p} = \mathbf{D}_{hex}^{p \to t} \mathbf{x}_{/\mathbf{B}_{hex}^p} \; \to \; \mathbf{x}'_{/\mathbf{B}_{hex}^t} = \mathbf{C}_{hex}^{t \to p} \mathbf{x}_{/\mathbf{B}_{hex}^p} \qquad (6)$$

### 1.2. Construction of the distortion, misorientation and correspondence matrices

The crystallographic features of a twin model are determined by the choice of a supercell. This supercell defines a sub-lattice of the hexagonal lattice; and it is actually this sub-lattice that is linearly distorted by $\mathbf{D}^{p \to t}$; the atoms inside the supercell do not follow the same trajectories as those at the corners of the cells; they "shuffle". The supercell is formed by three crystallographic directions **A**, **B**, **C** defining a matrix

$$\mathbf{B}_{super}^p = [\mathbf{B}_{hex}^p \to \mathbf{B}_{super}^p] = (\mathbf{A}, \mathbf{B}, \mathbf{C})_{/\mathbf{B}_{hex}^p}.$$

After distortion, the vectors of this basis are transformed into **A'**, **B'**, **C'** that define a new basis expressed in $\mathbf{B}_{hex}^{p}$ by the matrix $\mathbf{B}_{super}^{p'} = (\mathbf{A}', \mathbf{B}', \mathbf{C}')_{/\mathbf{B}_{hex}^{p}} = [\mathbf{B}_{hex}^{p} \rightarrow \mathbf{B}_{super}^{p'}]$. When the vectors are expressed in the $\mathbf{B}_{hex}^{t}$, it takes the form $\mathbf{B}_{super}^{t} = (\mathbf{A}', \mathbf{B}', \mathbf{C}')_{/\mathbf{B}_{hex}^{t}} = [\mathbf{B}_{hex}^{t} \rightarrow \mathbf{B}_{super}^{t}]$.

As $\mathbf{B}_{super}^{p'}$ and $\mathbf{B}_{super}^{t}$ express the same vectors, we get

$$[\mathbf{B}_{super}^{p'} \rightarrow \mathbf{B}_{super}^{t}] = \mathbf{I} \qquad (7)$$

with **I** the identity matrix. Building a crystallographic model dedicated to a specific twin consists in finding the appropriate vectors **A**, **B**, **C** of the supercell and finding how they are transformed into **A'**, **B'**, **C'**. The three important matrices previously defined can be calculated from the supercell.

The *distortion matrix* is expressed in $\mathbf{B}_{super}^{p}$ by

$$\mathbf{D}_{super}^{p \rightarrow t} = [\mathbf{B}_{super}^{p} \rightarrow \mathbf{B}_{super}^{p'}] = [\mathbf{B}_{super}^{p} \rightarrow \mathbf{B}_{hex}^{p}][\mathbf{B}_{hex}^{p} \rightarrow \mathbf{B}_{super}^{p'}] = (\mathbf{B}_{super}^{p})^{-1} \mathbf{B}_{super}^{p'}$$

As the distortion matrix is an active matrix; writing it in the basis $\mathbf{B}_{hex}^{p}$ leads to

$$\mathbf{D}_{hex}^{p \rightarrow t} = [\mathbf{B}_{hex}^{p} \rightarrow \mathbf{B}_{super}^{p}] \cdot \mathbf{D}_{super}^{p \rightarrow t} \cdot [\mathbf{B}_{super}^{p} \rightarrow \mathbf{B}_{hex}^{p}] = \mathbf{B}_{super}^{p} (\mathbf{B}_{super}^{p})^{-1} \mathbf{B}_{super}^{p'} (\mathbf{B}_{super}^{p})^{-1}, \text{ i.e.}$$

$$\mathbf{D}_{hex}^{p \rightarrow t} = \mathbf{B}_{super}^{p'} (\mathbf{B}_{super}^{p})^{-1} \qquad (8)$$

The *misorientation matrix* is $\mathbf{T}_{hex}^{p \rightarrow t} = [\mathbf{B}_{hex}^{p} \rightarrow \mathbf{B}_{hex}^{t}] = [\mathbf{B}_{hex}^{p} \rightarrow \mathbf{B}_{super}^{t}][\mathbf{B}_{super}^{t} \rightarrow \mathbf{B}_{hex}^{t}]$, i.e.

$$\mathbf{T}_{hex}^{p \rightarrow t} = \mathbf{B}_{super}^{p} (\mathbf{B}_{super}^{t})^{-1} \qquad (9)$$

The *correspondence matrix* is $\mathbf{C}_{hex}^{t \rightarrow p} = [\mathbf{B}_{hex}^{t} \rightarrow \mathbf{B}_{hex}^{p'}] = [\mathbf{B}_{hex}^{t} \rightarrow \mathbf{B}_{hex}^{p}][\mathbf{B}_{hex}^{p} \rightarrow \mathbf{B}_{hex}^{p'}]$, i.e. $\mathbf{C}_{hex}^{t \rightarrow p} = \mathbf{T}_{hex}^{t \rightarrow p} \mathbf{D}_{hex}^{p \rightarrow t}$, as found in equation (5). It can also be decomposed into

$$\mathbf{C}_{hex}^{t \rightarrow p} = [\mathbf{B}_{hex}^{t} \rightarrow \mathbf{B}_{super}^{t}][\mathbf{B}_{super}^{t} \rightarrow \mathbf{B}_{hex}^{p'}] = [\mathbf{B}_{hex}^{t} \rightarrow \mathbf{B}_{super}^{t}][\mathbf{B}_{super}^{t} \rightarrow \mathbf{B}_{super}^{p'}][\mathbf{B}_{super}^{p'} \rightarrow \mathbf{B}_{hex}^{p'}].$$

As the coordinates of the supercell are not changed by the distortion $[\mathbf{B}_{hex}^{p'} \rightarrow \mathbf{B}_{super}^{p'}] = [\mathbf{B}_{hex}^{p} \rightarrow \mathbf{B}_{super}^{p}] = \mathbf{B}_{super}^{p}$, and by using (7), we get

$$\mathbf{C}_{hex}^{t \rightarrow p} = \mathbf{B}_{super}^{t} (\mathbf{B}_{super}^{p})^{-1} \qquad (10)$$

As the matrices $\mathbf{B}_{super}^{p}$ and $\mathbf{B}_{super}^{t}$ are constituted by the crystallographic directions forming the supercell, their values are integers. As the inverse of an integer matrix is a rational matrix, the correspondence matrix is a rational matrix.

## 1.3. Obliquity correction

It is usual in the crystallographic models of ferroelectrics to introduce an obliquity correction. This is a rotation with a small angle (few degrees) that is composed with a stretch distortion matrix in order to transform it into a simple shear matrix. An obliquity correction can be introduced to correct a small tilt on a plane and/or an small rotation of a direction. Here we need to introduce a general obliquity correction function $\mathbf{Obl}(g, g', u, u')$. This function gives the rotation matrix noted $\mathbf{Obl}$ such that $\mathbf{Obl}(g) = g'$ and $(u) = u'$. Let us consider a direction $u$ and a plane $g$ expressed in the hexagonal basis. Expressed in the orthonormal basis $\mathbf{B}_{ortho}$ they are $u_o = \mathbf{H}_{hex} u$ and $g_o = \mathbf{H}^*_{hex} g$. In this basis the plane $g$ has the same coordinates as its normal direction $n_o$. A third direction defined by $l_0 = n_o \wedge u_o$ allows building another orthonormal basis $\mathbf{B}(g, u) = (\frac{u_o}{\|u_o\|}, \frac{n_o}{\|n_o\|}, \frac{l_0}{\|l_0\|})$. We build the orthonormal bases $\mathbf{B}(g, u)$ and $\mathbf{B}(g', u')$. The obliquity rotation is

$$\mathbf{Obl}(g, g', u, u') = \mathbf{B}(g', u')\big(\mathbf{B}(g, u)\big)^{-1} \tag{11}$$

It is a rotation matrix expressed in the orthonormal basis $\mathbf{B}_{ortho}$ that transforms $\mathbf{B}(g, u)$ into $\mathbf{B}(g', u')$. This rotation should be compensated by its inverse in order to put in coincidence the plane $g$ with the plane $g'$, and the direction $u$ with the direction $u'$.

## 1.4. Definition of unconventional twinning

We call *conventional* twin a twin whose lattice distortion is expressed by a simple shear matrix. The habit plane of these twins is the shear plane, which is also the plane maintained fully invariant by the shear distortion. This means that for two non-collinear directions $u$ and $v$ of the plane $g$, i.e. such that $g.u = g.v = 0$, are invariant by the distortion: $\mathbf{D}^{p \to t} u = u$ and $\mathbf{D}^{p \to t} v = v$. This implies that the dimension of the space formed by the kernel of $\mathbf{D}^{p \to t} - \mathbf{I}$ is such that

$$\text{Dim}(\text{Ker}(\mathbf{D}^{p \to t} - \mathbf{I})) = 2 \tag{12}$$

If the plane $g$ is invariant, it is untilted. Therefore, a consequence of the existence of an invariant plane is

$$g' = \big(\mathbf{D}^{p \to t}\big)^* g = \lambda\, g. \tag{13}$$

which means that $g$ is an eigenvector of $\big(\mathbf{D}^{p \to t}\big)^*$.

It should be noted that (12)⇒(13), but the reciprocal is not always true.

By noting the plane $g$ by its Miller indices $g = (h,k,l)$, and considering that the interplanar distance $d_{hkl} = \frac{1}{\|g\|}$, we get

$$\frac{1}{\lambda} = \frac{d'_{hkl}}{d_{hkl}} \tag{14}$$

As the plane is invariant, the volume change is completely given by $1/\lambda$. If $\lambda=1$, there is no volume change, the shear is called "simple shear". In the more general case, the shear is sometimes called

"invariant plane strain" (IPS) and not "shear" in order to distinguish it from pure shear (stretch). To our knowledge, all the deformation twins reported in literature till now are simple shear.

In the manuscript, we call *unconventional* twin a twin defined by a distortion matrix for which a plane is *untilted*, but *not invariant*. Mathematically it means that the distortion matrix checks equation (13) but not equation (12). The untilted plane is transformed into a plane that is not equivalent to the initial one by any of the crystal symmetries; some of the directions contained in the plane are modified in length and/or angle. To our knowledge, unconventional twinning has never been reported till now.

## 2. Unconventional $(212) \to (012)$ twinning mode built by obliquity correction of the (58°, a + 2b) prototype stretch twin

### 2.1. The (58°, a + 2b) prototype stretch twin

*The calculations were performed with Mathematica (see **Supplementary Data** Part A). This twin mode is also largely described in a separate paper[28]; only its main characteristics are recalled here.*

Let us use the letter "p" for the parent crystal, and "gr" for the ~(58°, **a**+2**b**) twins colored in "green" in the EBSD maps. The $p \to gr$ distortion is associated with the transformation $\mathbf{OX}_2 \to \mathbf{OX'}_2$, $\mathbf{OY} \to \mathbf{OY'}$, $\mathbf{OG} \to \mathbf{OG'}$ such that

- $\mathbf{OX}_2 = [200]_p$ is parallel to the twin vector $\mathbf{OX}_2' = [\bar{1}01]_{gr}$
- $\mathbf{OY}$ is invariant, $\mathbf{OY} = [120]_p$ is equal to $\mathbf{OY}' = [120]_{gr}$
- $\mathbf{OG} = [\bar{1}01]_p$ is parallel to the twin vector $\mathbf{OG}' = [\bar{2}00]_{gr}$

The vectors are here expressed by their hexagonal coordinates. The distortion matrix is thus[28]

$$\mathbf{F}_{hex}^{p \to gr} = \begin{pmatrix} \frac{\sqrt{1+\gamma^2}}{2} & \frac{2-\sqrt{1+\gamma^2}}{4} & \frac{\gamma^2-3}{2\sqrt{1+\gamma^2}} \\ 0 & 1 & 0 \\ 0 & 0 & \frac{2}{\sqrt{1+\gamma^2}} \end{pmatrix} \quad (15)$$

The values of the principal strains can be calculated in the cases of ideal hard-sphere packing and pure magnesium; they are (-4.2%, 0, +4.4%) or (-4.6%, 0, +4.8%), respectively.

The correspondence matrix is calculated by considering the vectors $\mathbf{OX}_2, \mathbf{OY}$, and $\mathbf{OG}$, and the vectors $\mathbf{OX}_2', \mathbf{OY}'$, and $\mathbf{OG}'$, in their respective hexagonal bases , i.e. by using the supercell $(\mathbf{OX}_2, \mathbf{OY}, \mathbf{OG})$:

$$\mathbf{B}_{super}^{p} = \begin{pmatrix} 2 & 1 & -1 \\ 0 & 2 & 0 \\ 0 & 0 & 1 \end{pmatrix} \text{ and } \mathbf{B}_{super}^{gr} = \begin{pmatrix} 1 & 1 & -2 \\ 0 & 2 & 0 \\ 1 & 0 & 0 \end{pmatrix} \quad (16)$$

The expressions of the correspondence matrix in the direct and reciprocal space are:

$$\mathbf{C}_{hex}^{gr \to p} = \mathbf{B}_{super}^{gr} \cdot \left(\mathbf{B}_{super}^{p}\right)^{-1} = \begin{pmatrix} \frac{1}{2} & \frac{1}{4} & -\frac{3}{2} \\ 0 & 1 & 0 \\ \frac{1}{2} & -\frac{1}{4} & \frac{1}{2} \end{pmatrix}, \text{ and} \quad (17)$$

$$\left(\mathbf{C}_{hex}^{gr \to p}\right)^{*} = \left(\mathbf{C}_{hex}^{gr \to p}\right)^{-T} = \begin{pmatrix} \frac{1}{2} & 0 & -\frac{1}{2} \\ \frac{1}{4} & 1 & \frac{1}{4} \\ \frac{3}{2} & 0 & \frac{1}{2} \end{pmatrix}$$

The misorientation matrix is given by equation (5):

$$\mathbf{T}_{hex}^{p \to t} = \mathbf{D}_{hex}^{p \to t}\left(\mathbf{C}_{hex}^{t \to p}\right)^{-1} = \begin{pmatrix} \frac{1}{\sqrt{1+\gamma^2}} & 0 & \frac{\gamma}{\sqrt{1+\gamma^2}} \\ 0 & 1 & 0 \\ -\frac{\gamma}{\sqrt{1+\gamma^2}} & 0 & \frac{1}{\sqrt{1+\gamma^2}} \end{pmatrix} \quad (18)$$

which is a rotation of angle $ArcCos\left(\frac{1}{\sqrt{1+\gamma^2}}\right) = ArcTan(\gamma)$, that is equal to 58.5° for hard-sphere packing and 58.4° for magnesium.

Some correspondences between some planes and directions of the parent and its twins calculated from the correspondence matrices in equation (17) are interesting to interpret the EBSD map. They are given in Table 1.

| Parent | | → | Twin | |
|---|---|---|---|---|
| *Planes* | | | | |
| **(212)** | ∈ {12$\bar{3}$2} | → | **(024)** | ∈ {02$\bar{2}$4} |
| (004) | ∈ {0004} | → | ($\bar{2}$12) | ∈ {$\bar{2}$112} |
| *Directions* | | | | |
| **[0$\bar{2}$1]** | ∈ $\frac{1}{3}$⟨22$\bar{4}$3⟩ | → | **[$\bar{2}\bar{2}$1]** | ∈ $\frac{1}{3}$⟨22$\bar{4}$3⟩ |
| [200] | ∈ $\frac{1}{3}$⟨22$\bar{4}$0⟩ | → | [101] | ∈ $\frac{1}{3}$⟨11$\bar{2}$3⟩ |
| [$\bar{1}$01] | ∈ $\frac{1}{3}$⟨11$\bar{2}$3⟩ | → | [$\bar{2}$00] | ∈ $\frac{1}{3}$⟨22$\bar{4}$0⟩ |
| [120] | ∈ ⟨01$\bar{1}$0⟩ | → | [120] | ∈ ⟨01$\bar{1}$0⟩ |

*Supplementary Table 1. Correspondence between some planes and between some directions established by the (58°,**a**) stretch twin. The families of their equivalent directions/planes are indicated by using the four-index Miller-Bravais notations. The plane $\mathbf{g}_0 = (212)_p$ and the direction $\mathbf{u}_0 = [0\bar{2}1]_p$ of the parent crystal (in bold) will be used to build the model of the green twins.*

From this table, we tried two different approaches to build a model that could explain the green twins observed experimental EBSD maps. The first approach was the most intuitive one; it is based on the fact that the direction $\mathbf{OY} = [120]_p$ is invariant in the stretch twin model (see Supplementary Table 1). However, after many attempts, this way was given up because all the habit planes we could predict contain the **OY** direction, which is not in agreement with the observations. A dissymmetry should be introduced in the system. The second approach was less intuitive; but it was revealed to fit perfectly with the observations, even for small details that were not noticed at the beginning. It is based on the correspondence between the $(212)_p$ and $(012)_{gr}$ planes, and between the $[0\bar{2}1]_p$ and $[\bar{2}\bar{2}1]_{gr}$ directions (Supplementary Table 1). The model, described in the next section, introduces an obliquity correction such that the plane $\boldsymbol{g_0} = (212)_p$ becomes untilted and the direction $\boldsymbol{u_0} = [0\bar{2}1]_p$ invariant.

### 2.2. Unconventional twin derived from the (58°, a + 2b) stretch twin prototype

*The calculations were performed with Mathematica (see **Supplementary Data** Part B).*

The EBSD map shows that the habit plane of the green twin is not invariant; it is the plane $(212)_p$ transformed into the plane $(012)_{gr}$. These two planes are not equivalent. The modification of this plane comes from the transformation of the directions it contains, i.e. $[\bar{1}01]_p$ is transformed into $[\bar{2}00]_{gr}$, and $[0\bar{2}1]_p$ is transformed into $[\bar{2}\bar{2}1]_{gr}$. The transformation of the direction $[\bar{1}01]_p$ occurs by a stretch of $\frac{2}{\sqrt{1+\gamma^2}} \approx 1.04$. There is no stretch for the $[0\bar{2}1]_p$ direction because it is equivalent to $[\bar{2}\bar{2}1]_{gr}$. In addition, the angle formed by the pairs $\alpha_p = ([\bar{1}01]_p, [0\bar{2}1]_p) = ArcCos(\frac{\gamma^2-1}{\sqrt{4+5\gamma^2+\gamma^4}}) \approx 70.56°$ is slightly reduced to become that the angle between the pair $\alpha_t = ([0\bar{2}1]_{gr}, [\bar{2}\bar{2}1]_{gr}) = ArcCos(\frac{1}{\sqrt{4+\gamma^2}}) \approx 67.15°$. The stretch of the $[\bar{1}01]_p$ direction (+4%) and the angular distortion of the plane (-3°) are quite small. We noticed that the planar transformation $(212)_p \rightarrow (012)_{gr}$ can be explained by the displacements of the atoms located in the upper layer l = 1/3 of the plane $(212)_p$ as described in **Fig.6**.

Even if the direction $\boldsymbol{u_0} = [0\bar{2}1]_p$ is not stretched, the prototype twin induces a slight rotation of angle $\xi_u$ of this direction. This is this rotation that should be compensated in order to build the model. The rotation angle $\xi_u$ can be calculated by working in the orthonormal basis; it is the angle between $\mathbf{H}_{hex}\boldsymbol{u_0}$ and $\mathbf{F}_{ortho}^{p \rightarrow t} \cdot \mathbf{H}_{hex}\boldsymbol{u_0}$, with $\mathbf{F}_{ortho}^{p \rightarrow t}$ the inverse of the transpose of the matrix (15). The calculations show that

$$\xi_u = ArcCos\left(\frac{-1 + 3\gamma^2 + 3\sqrt{1+\gamma^2}}{\sqrt{1+\gamma^2}(4+\gamma^2)}\right) \tag{19}$$

For a hard-sphere packing ratio $\gamma = \sqrt{\frac{8}{3}}$, the obliquity is $\xi_u = ArcCos\left(\frac{99+21\sqrt{33}}{220}\right) \approx 3.29°$.

The stretch prototype twin also induces a rotation of the plane $\boldsymbol{g_0} = (212)_p$. The rotation angle $\xi_g$ is the angle between $\mathbf{H}_{hex}^* \boldsymbol{g_0}$ and $\left(\mathbf{F}_{ortho}^{p \rightarrow t}\right)^* \mathbf{H}_{hex}^* \boldsymbol{g_0}$, with $\left(\mathbf{F}_{ortho}^{p \rightarrow t}\right)^*$ the inverse of the transpose of the matrix (15). The calculations show that

$$\xi_g = ArcCos\left(\frac{3 + \gamma^2(3 + 2\sqrt{1+\gamma^2})}{\sqrt{(1+\gamma^2)(3+\gamma^2)(3+7\gamma^2)}}\right) \tag{20}$$

For a hard-sphere packing ratio $\gamma = \sqrt{\frac{8}{3}}$, the obliquity is $\xi_g = ArcCos\left(\frac{16+3\sqrt{33}}{\sqrt{1105}}\right) \approx 1.24°$.

In order to correct in one shot the two obliquities $\xi_u$ and $\xi_g$ and rotate the stretch prototype twin such that the direction $\boldsymbol{u_0} = [0\bar{2}1]_p$ becomes invariant and the plane $\boldsymbol{g_0} = (212)_p$ becomes untilted, the general obliquity correction function $\mathbf{Obl}(\boldsymbol{g}, \boldsymbol{g'}, \boldsymbol{u}, \boldsymbol{u'})$ described in section 1.3 is used with $\boldsymbol{g} = (212)_p$, $\boldsymbol{g'} = \left(\mathbf{F}_{hex}^{p \to t}\right)^* \boldsymbol{g}$, $\boldsymbol{u} = [0\bar{2}1]_p$, $\boldsymbol{u'} = \mathbf{F}_{hex}^{p \to gr} \boldsymbol{u}$. The result expressed as a function of $\gamma$ is too long to be written here, even by writing separately each of its nine components. The reader can however see the result in Part B of the Mathematica program in **Supplementary Data**.

In the special case of a hard-sphere packing ratio $\gamma = \sqrt{\frac{8}{3}}$, the approximate numerical value of the obliquity matrix is:

$$\mathbf{O}_{g,u} = \mathbf{Obl}(\boldsymbol{g}, \boldsymbol{g'}, \boldsymbol{u}, \boldsymbol{u'}) \approx \begin{pmatrix} 0.9764 & 0.0452 & -0.0392 \\ -0.0443 & 1.0212 & 0.0423 \\ 0.0231 & -0.0227 & 0.9991 \end{pmatrix} \quad (21)$$

The obliquity angle is $ArcCos\left(\frac{1}{440}(-121 + 21\sqrt{33}) + \frac{803 + 369\sqrt{33}}{88\sqrt{1105}}\right) \approx 3.33°$

The distortion matrix corrected of the obliquity is

$$\mathbf{D}_{hex}^{p \to gr} = \mathbf{O}_{g,u}^{-1} \cdot \mathbf{F}_{hex}^{p \to gr} \quad (22)$$

Despite the very long analytical expression of the general form of the obliquity matrix $\mathbf{R}$, the distortion corrected from this obliquity can be calculated and simplified. The analytical expressions of the nine components $\mathbf{D}_{ij}$ of $\mathbf{D}_{hex}^{p \to gr}$ are:

$$\mathbf{D}_{11} = \frac{A}{6 + 2\gamma^2} \quad (23)$$

$$\mathbf{D}_{12} = \frac{3(\gamma^2 - 3)}{4A}$$

$$\mathbf{D}_{13} = \frac{3(\gamma^2 - 3)}{2A}$$

$$\mathbf{D}_{21} = \frac{3 - 2\gamma^2 - \gamma^4 + A}{12 + 7\gamma^2 + \gamma^4}$$

$$\mathbf{D}_{22} = \frac{-9 + 4\gamma^4 + 5A + \gamma^2(15 + A)}{2(4 + \gamma^2)A}$$

$$\mathbf{D}_{23} = \frac{4\gamma^4 - \gamma^2(-15 + A) - 3(3 + A)}{(4 + \gamma^2)A}$$

$$\mathbf{D}_{31} = \frac{-3 + 2\gamma^2 + \gamma^4 - A}{2(12 + 7\gamma^2 + \gamma^4)}$$

$$\mathbf{D}_{32} = \frac{-45 - 59\gamma^4 - 7\gamma^6 + 57A + \gamma^2(-129 + 13A)}{4(36 + 105\gamma^2 + 52\gamma^4 + 7\gamma^6)}$$

$$\mathbf{D}_{33} = \frac{27 + 45\gamma^4 + 7\gamma^6 + 57B + \gamma^2(81 + 13A)}{2(36 + 105\gamma^2 + 52\gamma^4 + 7\gamma^6)}$$

with $A = \sqrt{(3+\gamma^2)(3+7\gamma^2)}$

It is checked that this distortion matrix $\mathbf{D}_{hex}^{p \to gr}$ maintains invariant the direction $\mathbf{u_0} = [0\bar{2}1]_p$ and that $\left(\mathbf{D}_{hex}^{p \to t}\right)^*$ maintains untilted the plane $\mathbf{g_0} = (212)_p$.

For the ideal hard-sphere packing ratio, the distortion matrix takes the value

$$\mathbf{D}_{hex}^{p \to gr} = \begin{pmatrix} \frac{\sqrt{\frac{65}{17}}}{2} & -\frac{3}{4\sqrt{1105}} & -\frac{3}{2\sqrt{1105}} \\ -\frac{1}{4} + \frac{3\sqrt{\frac{13}{85}}}{4} & \frac{23}{40} + \frac{107}{8\sqrt{1105}} & -\frac{17}{20} + \frac{107}{4\sqrt{1105}} \\ \frac{1}{8} - \frac{3\sqrt{\frac{13}{85}}}{8} & -\frac{23}{80} + \frac{33\sqrt{\frac{5}{221}}}{16} & \frac{17}{40} + \frac{33\sqrt{\frac{5}{221}}}{8} \end{pmatrix} \quad (24)$$

$$\approx \begin{pmatrix} 0.9777 & -0.0226 & -0.0451 \\ 0.0433 & 0.9774 & -0.0453 \\ -0.0217 & 0.0227 & 1.0455 \end{pmatrix}$$

As $\mathbf{D}_{hex}^{p \to gr}$ differs from $\mathbf{F}_{hex}^{p \to gr}$ only by the obliquity correction, the correspondence matrix given by equation (17) is not affected. The distortion $\mathbf{D}_{hex}^{p \to gr}$ is *unconventional* as the untilted plane $(212)_p$, which is also the habit plane of the green twin, is not fully invariant but transformed into the plane $(012)_{gr}$. The modes of plasticity required to accommodate this deformation are not the subject of the paper, but it is hoped that deeper TEM investigations and molecular dynamics simulations can bring important elements of responses.

We have seen in section 2.1 that the rotation matrix $\mathbf{T}_{hex}^{p \to gr}$ between the parent and the green twin associated with the stretch distortion $\mathbf{F}_{hex}^{p \to gr}$ is a rotation of axis $\mathbf{OY} = [120]_{hex}$ and of angle $\theta_F = ArcCos(\frac{1}{\sqrt{1+\gamma^2}})$, that is 58.5° for hard-sphere packing, 58.4° for magnesium. Now, if instead of the stretch prototype $\mathbf{F}_{hex}^{p \to gr}$, the distortion $\mathbf{D}_{hex}^{p \to gr}$ applies, the orientation of the twinned crystal is slightly modified. The new expression of orientation matrix $\mathbf{T}_{hex}^{p \to gr}$ between the twin and the parent is obtained by using the distortion matrix and the correspondence matrix in equation (5); it is calculated in part B of **Supplementary Data**:

$$\mathbf{T}_{hex}^{p \to gr} = \begin{pmatrix} \sqrt{\frac{3+\gamma^2}{3+7\gamma^2}} & \frac{-3+2\gamma^2}{A} & \frac{6\gamma^2}{A} \\ \frac{2(3-\gamma^2(2+\gamma^2)+A)}{(4+\gamma^2)A} & \frac{-6+13\gamma^2+3\gamma^4+2A}{(4+\gamma^2)A} & \frac{2\gamma^2(9+\gamma^2-A)}{(4+\gamma^2)A} \\ \frac{18}{3+7\gamma^2-5A} & \frac{9(12+\gamma^2)}{9+7\gamma^4+21A+4\gamma^2(6+A)} & \frac{24\gamma^4+7\gamma^6+12A+\gamma^2(9-2A)}{(3+7\gamma^2)(12+7\gamma^2+\gamma^4)} \end{pmatrix} \quad (25)$$

with $A = \sqrt{(3+\gamma^2)(3+7\gamma^2)}$

The rotation angle is

$$\theta_D = ArcCos\left(\frac{9 + 9\gamma^2 + 2\gamma^4 - A}{(4 + \gamma^2)A}\right) \qquad (26)$$

The rotation axis is a complex form of the packing ratio $\gamma$; it slightly deviates from the axis **OY** = $[120]_{hex}$. In the case of the ideal hard-sphere packing ratio, the rotation angle is $\theta_D = ArcCos\left(-\frac{3}{20} + \frac{\sqrt{\frac{85}{13}}}{4}\right) \approx 60.71°$, and the axis is $\left[1, 2, \frac{3}{(31+\sqrt{1105})}\right]_{hex} \approx [1, 2, 0.047]_{hex}$. For magnesium the angle is $\theta_D \approx 60.76°$, and the axis is $\approx [1, 2, 0.051]_{hex}$. Consequently, a careful examination of the rotation angle of the misorientation between the twin and its parent permits to know whether this twin directly comes from the prototype stretch distortion $\mathbf{F}_{hex}^{p \to gr}$ or from its derived obliquity-corrected form, i.e. $\mathbf{D}_{hex}^{p \to gr}$. In the former case the misorientation angle is close 58° and in the latter case is close to 61°. Both forms exist in the EBSD map of **Fig.1**.

The other method to distinguish the twin generated by distortion $\mathbf{F}_{hex}^{p \to gr}$ from the one generated by $\mathbf{D}_{hex}^{p \to gr}$ consists in considering the $\langle 201 \rangle$ directions. All are rotated by the distortion $\mathbf{F}_{hex}^{p \to gr}$ whereas the distortion matrix $\mathbf{D}_{hex}^{p \to gr}$ maintains the direction $\boldsymbol{u_0} = [0\bar{2}1]_p$ invariant.

A rotation equivalent to $\mathbf{T}_{hex}^{p \to gr}$ that has $\boldsymbol{u_0}$ for rotation axis is found by using a 6-fold rotation symmetry. In the basis $\mathbf{B}_{hex}$, and noted by its Seitz symbol, this symmetry is

$$\mathbf{6}_{001}^- = \begin{pmatrix} 0 & 1 & 0 \\ -1 & 1 & 0 \\ 0 & 0 & 1 \end{pmatrix} \qquad (27)$$

The equivalent rotation between the twin and the parent crystal is $\mathbf{T}_{hex}^{p \to gr} \cdot \mathbf{6}_{001}^-$. This rotation matrix is explicitly written in PartB of **Supplementary Data**. The rotation angle is

$$\theta_D = ArcCos\left(\frac{6 - \gamma^2}{2A}\right) \qquad (28)$$

In the case of hard-sphere packing, this angle is $\theta_D = ArcCos\left(\sqrt{\frac{5}{221}}\right) \approx 81.35°$. It was checked that the rotation axis is indeed $\boldsymbol{u_0} = [0\bar{2}1]_p$, independently of the packing ratio $\gamma$. The commercial EBSD programs do not give all the equivalent rotations but only the disorientation, i.e. the rotation that among all the equivalent rotations has the lowest angle. The present example shows that this choice is sometimes not well adapted, as the rotation axis of $\mathbf{T}_{hex}^{p \to gr}$ is complex, even if close to [120], whereas that of $\mathbf{T}_{hex}^{p \to gr} \cdot \mathbf{6}_{001}^-$ is simply a rotation around the $[0\bar{2}1]$ direction.

## 3. Unconventional $(012) \to (212)$ twinning mode by obliquity correction of the (86°, a) twin

The experimental EBSD maps show that the extension "yellow" twins are often co-formed with the "green" twins and constitute green-yellow "stripes" as that in the green rectangle of **Fig.1a**. In the EBSD map acquired in the cross-section B, the yellow twins can also appear orange or red, as shown in the **Extended Data Fig.1**. The striking point is that these the "yellow" twins are conventional twins of the parent "grey" crystal: their habit plane is the plane $(212)_p$, and this plane is common to both the parent and "yellow" crystal. The misorientation between the "yellow" twins and the parent "grey" crystal experimentally measured from the EBSD maps is a rotation of 48° around an axis close to a ⟨241⟩ direction, as shown in **Fig.1b and c**. To the best of our knowledge this twin has never been reported or predicted; which means that, even if conventional, there is not yet crystallographic model for it. In order to build such a model, additional information is required. We noticed that the misorientation between the yellow twins and the green twins is close to (86°, **a**), with an interface plane close to {102}, which means that the yellow and green twins are linked by an kind of extension twin relation, or a twinning relation close to that one.

The crystallographic model of (86°, **a**) extension twinning in hcp metals was proposed by correcting the obliquity of a (90°, **a**) prototype stretch twin to maintain a plane {102} untilted[21]. The correspondence matrix written in the reciprocal space shows that among the five other equivalent {102} planes, one is also transformed into another {102} plane (by conjugation), and the four other ones are transformed into {212} planes. Some of these four {102} planes transformed into {212} planes are only slightly tilted during the extension twinning. Thus, it is possible, by adding an obliquity correction to a conventional extension twin, to change the conventional extension twin into an unconventional twin that transforms a {102} plane into a {212} plane without tilt. The green twin transforms a $\{212\}_p$ plane into a $\{102\}_{gr}$ plane, and the yellow twin would transform back this $\{102\}_{gr}$ plane into a $\{212\}_p$ plane, such that the yellow twin would leave invariant the $\{212\}_p$ plane of the parent crystal, i.e. $\{212\}_p = \{212\}_{ye}$, as observed in the EBSD maps. Before detailing the obliquity correction that will be applied to the conventional extension twin, let us determine the appropriate reference frame that should be used to express the extension twinning distortion matrix in order to be composed with the green twin.

### 3.1. The conventional (86°, a) twin in an adequate basis

*The calculations were performed with Mathematica (see **Supplementary Data** Part C).*

In order to build the unconventional yellow twin derived from a conventional (86°, ***a***) extension twin, we have to quickly recall some the crystallographic details of this twin. The (86°, ***a***) extension twin described in the paper[21] is an extension twin on the plane $(0\bar{1}2)_p$. This twin was shown to derive from a stretch prototype, called (90°, ***a***) twin. Most of the calculations[21] were done by assuming an ideal hard-sphere packing ratio in order to determine the continuous form of the distortion. The calculations related to the general case depending on $\gamma$ were not explicitly detailed. Let us present them now. The distortion matrix associated with the (90°, ***a***) twin is[21]

$$\mathbf{U}_{hex}^{p \to t} = \begin{pmatrix} 1 & 0 & 0 \\ 0 & \frac{\gamma}{\sqrt{3}} & 0 \\ 0 & 0 & \frac{\sqrt{3}}{\gamma} \end{pmatrix} \qquad (29)$$

This (90°, ***a***) stretch prototype twin induces a rotation of the plane $\mathbf{g} = (0\bar{1}2)_p$ around the axis $[100]_p$. This rotation $\mathbf{R}_g$ has for rotation angle $\xi_g$ that can be calculated by working in the orthonormal basis; it is the angle between $\mathbf{H}_{hex}^* \mathbf{g}$ and $\left(\mathbf{U}_{ortho}^{p \to t}\right)^* . \mathbf{H}_{hex}^* \mathbf{g}$, with $\left(\mathbf{U}_{ortho}^{p \to t}\right)^*$ the inverse of the transpose of the matrix (29). The calculations computed in Part C of **Supplementary Data** show that

$$\xi_g = ArcCos\left(\frac{2\sqrt{3}\gamma}{3+\gamma^2}\right) \tag{30}$$

After the obliquity correction, the distortion matrix becomes

$$\mathbf{E}^{p\to t}_{hex} = \mathbf{R}_g^{-1}\cdot\mathbf{U}^{p\to t}_{hex} = \begin{pmatrix} 1 & -\frac{3-\gamma^2}{2(3+\gamma^2)} & \frac{3-\gamma^2}{3+\gamma^2} \\ 0 & \frac{2\gamma^2}{3+\gamma^2} & \frac{2(3-\gamma^2)}{3+\gamma^2} \\ 0 & -\frac{3-\gamma^2}{2(3+\gamma^2)} & \frac{6}{3+\gamma^2} \end{pmatrix} \tag{31}$$

For the ideal hard-sphere packing ratio, the obliquity is $\xi_g = ArcCos\left(\frac{12\sqrt{2}}{17}\right) \approx 3.37°$, and the obliquity-corrected distortion matrix becomes

$$\mathbf{E}^{p\to t}_{hex} = \begin{pmatrix} 1 & -\frac{1}{34} & \frac{1}{17} \\ 0 & \frac{16}{17} & \frac{2}{17} \\ 0 & -\frac{1}{34} & \frac{18}{17} \end{pmatrix} \tag{32}$$

The distortion matrix (32) generates the conventional (86°, **a**) twin for which the invariant plane is $(0\bar{1}2)_p$. In order to continue working with coherent coordinates in the system formed by the "green", "yellow" and "grey" crystals, we need to use an extension twin such that, once combined with the green twin distortion (23), it yields a conventional twin on the $(212)_p$ plane. A hexagonal symmetry is thus introduced; its choice will be justified *a posteriori* by the internal coherency of the calculations and by the perfect agreement with the experimental EBSD observations. This internal symmetry noted by its Seitz symbol is

$$\mathbf{2}_{110} = \begin{pmatrix} 0 & 1 & 0 \\ 1 & 0 & 0 \\ 0 & 0 & -1 \end{pmatrix} \tag{33}$$

It allows establishing the distortion matrix of the $(102)_p$ extension twin from that of the $(0\bar{1}2)_p$ extension twin given in equation (32):

$$\mathbf{E}^{gr\to ye}_{hex} = (\mathbf{2}_{110})^{-1}\mathbf{E}^{p\to t}_{hex}\mathbf{2}_{110} = \begin{pmatrix} \frac{2\gamma^2}{3+\gamma^2} & 0 & -\frac{2(3-\gamma^2)}{3+\gamma^2} \\ -\frac{3-\gamma^2}{2(3+\gamma^2)} & 1 & -\frac{3-\gamma^2}{(3+\gamma^2)} \\ \frac{3-\gamma^2}{2(3+\gamma^2)} & 0 & \frac{6}{3+\gamma^2} \end{pmatrix} \tag{34}$$

To be clearer, we have used in equation (34) a notation that specifies that the parent crystal is the green grain and that the yellow grains are linked to it by an extension twin (even if not yet corrected by the obliquity). Indeed, the parent index "p" is here "gr" and the twin index "t" is "gr".

The correspondence matrices in the direct and reciprocal spaces are

$$\mathbf{C}_{hex}^{ye \to gr} = \mathbf{C}_{hex}^{ye \to gr} \mathbf{2}_{110} = \begin{pmatrix} -\frac{1}{2} & 1 & 1 \\ 0 & 0 & 2 \\ \frac{1}{2} & 0 & 0 \end{pmatrix} \tag{35}$$

$$\text{and } \left(\mathbf{C}_{hex}^{ye \to gr}\right)^* = \begin{pmatrix} 0 & 1 & 0 \\ 0 & -\frac{1}{2} & \frac{1}{2} \\ 2 & 1 & 0 \end{pmatrix}$$

And the misorientation matrix is

$$\mathbf{T}_{hex}^{gr \to ye} = (\mathbf{2}_{110})^{-1} \mathbf{T}_{hex}^{gr \to ye} = \begin{pmatrix} 0 & 1 - \dfrac{6}{3+\gamma^2} & \dfrac{4\gamma^2}{3+\gamma^2} \\ 1 & -\dfrac{3}{3+\gamma^2} & \dfrac{2\gamma^2}{3+\gamma^2} \\ 0 & \dfrac{3}{3+\gamma^2} & -1 + \dfrac{6}{3+\gamma^2} \end{pmatrix} \tag{36}$$

In the case of hard-sphere packing the distortion and orientation matrices take rational values:

$$\mathbf{E}_{hex}^{gr \to ye} = \begin{pmatrix} \dfrac{16}{17} & 0 & -\dfrac{2}{17} \\ -\dfrac{1}{34} & 1 & -\dfrac{1}{17} \\ \dfrac{1}{34} & 0 & \dfrac{18}{17} \end{pmatrix} \text{ and } \mathbf{T}_{hex}^{gr \to ye} = \begin{pmatrix} 0 & -\dfrac{1}{17} & \dfrac{32}{17} \\ 1 & -\dfrac{9}{17} & \dfrac{16}{17} \\ 0 & \dfrac{9}{17} & \dfrac{1}{17} \end{pmatrix} \tag{37}$$

Now that the appropriate basis is found to express the conventional extension twin, the additional obliquity correction required to get the planar distortion $(012) \to (212)$ without tilt can be determined.

### 3.2. The unconventional twin derived from the (86°, a) twin prototype

*The calculations were performed with Mathematica (see **Supplementary Data** Part D).*

The extension twin (34) leaves invariant the plane $(102)_{gr}$ and the direction $[\bar{2}\bar{2}1]_{gr}$, and it transforms the plane $(012)_{gr}$ into the plane $(212)_{ye}$ by the correspondence matrix (35), but this plane is tilted. Now, we will build by obliquity correction of the conventional extension twin (34) an unconventional twin such that the plane $(012)_{gr}$ is transformed into the plane $(212)_{ye}$ without tilt, and such that the direction $[\bar{2}\bar{2}1]_{gr}$ becomes invariant. This twin, when composed with the unconventional "green" twin, will give a conventional twin relatively to the "grey" parent crystal. In order to determine the obliquity matrix, one could directly apply the general function (21), but we noticed that correcting the obliquity of the plane $\mathbf{g} = (012)_{gr}$ is sufficient to also correct the obliquity of the direction $[\bar{2}\bar{2}1]_{gr}$, as detailed as follows.

The tilt $\xi_g$ of the plane $\mathbf{g} = (012)_{gr}$ by the conventional distortion matrix $\mathbf{E}_{hex}^{gr \to ye}$ can be calculated by working in the orthonormal basis; it is the angle between $\mathbf{H}_{hex}^* \mathbf{g}$ and $\mathbf{H}_{hex}^* \left(\mathbf{E}_{hex}^{gr \to ye}\right)^* \mathbf{g}$, with $\left(\mathbf{E}_{hex}^{gr \to ye}\right)^*$ the inverse of the transpose of the matrix (34):

$$\xi_g = ArcCos\left(\frac{18 + 27\gamma^2 + 5\gamma^4}{2(3 + \gamma^2)A}\right) \tag{38}$$

The obliquity rotation axis written in the hexagonal basis is

$$\boldsymbol{\omega}_g = \frac{3 - \gamma^2}{2\sqrt{2}(3 + \gamma^2)}[\bar{2}, \bar{2}, 1] \tag{39}$$

The rotation matrix required to compensate the tilt of the plane $\mathbf{g}_0 = (012)_p$ can thus be calculated, but its analytical expression depending on the packing ratio is too large to fit the page width. In the case of ideal $\gamma$ ratio, the obliquity rotation angle is $\xi_g = ArcCos\left(\frac{113}{17}\sqrt{\frac{5}{221}}\right) \approx 1.11°$ and the rotation axis in the hexagonal basis is $\boldsymbol{\omega}_g = \frac{1}{17\sqrt{2}}[\bar{2}, \bar{2}, 1]$.

The obliquity-corrected distortion matrix is noted $\mathbf{D}_{hex}^{gr \to ye}$. The analytical expressions of the nine components $\mathbf{D}_{ij}$ of $\mathbf{D}_{hex}^{gr \to ye}$ depending on the stacking ratio calculated with Mathematica are:

$$\mathbf{D}_{11} = \frac{2\left(-1 + \sqrt{\frac{3 + 7\gamma^2}{3 + \gamma^2}}\right) + \gamma^2\left(2 + \sqrt{\frac{3 + 7\gamma^2}{3 + \gamma^2}}\right)}{2(4 + \gamma^2)} \tag{40}$$

$$\mathbf{D}_{12} = \frac{-6 - 3\gamma^2 - \gamma^4 + 2A}{(4 + \gamma^2)A}$$

$$\mathbf{D}_{13} = \frac{11\gamma^2 + 5\gamma^4 - 6(1 + A)}{(4 + \gamma^2)A}$$

$$\mathbf{D}_{21} = \frac{-3 + 2\gamma^2 + \gamma^4 - A}{12 + 7\gamma^2 + \gamma^4}$$

$$\mathbf{D}_{22} = \frac{2(3 + 6\gamma^2 + \gamma^4 + A)}{(4 + \gamma^2)A}$$

$$\mathbf{D}_{23} = \frac{2(3 + 5\gamma^2 + 2\gamma^4 - 3A)}{(4 + \gamma^2)A}$$

$$\mathbf{D}_{31} = \frac{3 - 2\gamma^2 - \gamma^4 + A}{24 + 14\gamma^2 + 2\gamma^4}$$

$$\mathbf{D}_{32} = \frac{9 + \gamma^2 - A}{(4 + \gamma^2)A}$$

$$\mathbf{D}_{33} = \frac{3(9 + 7\gamma^4 + 7A + 3\gamma^2(8 + A))}{(4 + \gamma^2)A^2}$$

with $A = \sqrt{(3 + \gamma^2)(3 + 7\gamma^2)}$

In the case of ideal hard-sphere packing the distortion matrix takes the value:

$$\mathbf{D}_{hex}^{gr \to ye} = \begin{pmatrix} \dfrac{1}{4} + \dfrac{7\sqrt{\dfrac{13}{85}}}{4} & \dfrac{3}{10} - \dfrac{19}{2\sqrt{1105}} & -\dfrac{9}{10} + \dfrac{53}{2\sqrt{1105}} \\ \dfrac{1}{4} - \dfrac{3\sqrt{\dfrac{13}{85}}}{4} & \dfrac{3}{10} + \dfrac{47}{2\sqrt{1105}} & -\dfrac{9}{10} + \dfrac{11\sqrt{\dfrac{5}{221}}}{2} \\ -\dfrac{1}{8} + \dfrac{3\sqrt{\dfrac{13}{85}}}{8} & -\dfrac{3}{20} + \dfrac{21}{4\sqrt{1105}} & \dfrac{9}{20} + \dfrac{81}{4\sqrt{1105}} \end{pmatrix} \quad (41)$$

$$\approx \begin{pmatrix} 0.9344 & 0.0142 & -0.1028 \\ -0.0433 & 1.0069 & -0.0727 \\ 0.0217 & 0.0079 & 1.0592 \end{pmatrix}$$

It can checked that $\mathbf{D}_{hex}^{gr \to ye}$ leaves invariant the direction $\mathbf{u}_0 = [\bar{2}, \bar{2}, 1]_{gr}$ and leaves untilted the plane $(012)_{gr}$

The orientation of the unconventional twin is given by the misorientation matrix between the hexagonal bases. It is $\mathbf{T}_{hex}^{gr \to ye} = \mathbf{D}_{hex}^{gr \to ye} \left(\mathbf{C}_{hex}^{ye \to gr}\right)^{-1}$

$$\mathbf{T}_{hex}^{gr \to ye} \quad (42)$$
$$= \begin{pmatrix} \dfrac{-6 - 3\gamma^2 - \gamma^4 + 2A}{(4 + \gamma^2)A} & \dfrac{2(3 - \gamma^2(2 + \gamma^2) + A)}{(4 + \gamma^2)A} & -\dfrac{2\gamma^2(7 + 3\gamma^2 + A)}{(4 + \gamma^2)A} \\ \dfrac{2(3 + 6\gamma^2 + \gamma^4 + A)}{(4 + \gamma^2)A} & \dfrac{-6 - 11\gamma^2 - 3\gamma^4 + 2A}{(4 + \gamma^2)A} & -\dfrac{2\gamma^2(-1 + \gamma^2 + A)}{(4 + \gamma^2)A} \\ \dfrac{9 + \gamma^2 - A}{(4 + \gamma^2)A} & \dfrac{18}{3 + 7\gamma^2 - 5A} & \dfrac{-12 + \gamma^2(-8 + A)}{(4 + \gamma^2)A} \end{pmatrix}$$

with $A = \sqrt{(3 + \gamma^2)(3 + 7\gamma^2)}$

This orientation matrix is fully equivalent by internal symmetry to the matrix $\mathbf{T}_{hex}^{gr \to ye} \, \mathbf{2}_{100}$, which is a rotation around the axis $\mathbf{u}_0 = [\bar{2}, \bar{2}, 1]_{gr}$ and of angle

$$\theta_D = ArcCos\left(\dfrac{-3 - 2\gamma^2}{A}\right) \quad (43)$$

In the case of hard-sphere packing, this angle is $\theta_D = ArcCos\left(-5\sqrt{\dfrac{5}{221}}\right) \approx 138.77°$.

# 4. New conventional twin generated by the composition of the unconventional twins derived from the (58°, a + 2b) and (86°,b) twin prototypes.

*The calculations were performed with Mathematica (see **Supplementary Data** Part E).*

From the previous calculations and from the EBSD results, the crystallographic link between the parent, green and yellow twins can be summarized as follows:

1. The green twin results from an unconventional $(212)_p \to (012)_{gr}$ twinning of the parent crystal. The correspondence, distortion and orientation matrices associated with this twin mode are $\mathbf{C}_{hex}^{gr \to p}$, $\mathbf{D}_{hex}^{p \to gr}$ and $\mathbf{T}_{hex}^{p \to gr}$ given by equations (17), (23), and (25).
2. The yellow twins are linked to the green twins by an unconventional $(012)_{gr} \to (212)_{ye}$ twinning relationship that is an obliquity-corrected form of extension twinning. The correspondence, distortion and orientation matrices associated with this twin mode are $\mathbf{C}_{hex}^{ye \to gr}$, $\mathbf{D}_{hex}^{gr \to ye}$ and $\mathbf{T}_{hex}^{gr \to ye}$ given by equations (35), (40), and (42), respectively.
3. The yellow twins formed by the combination of the two unconventional twins, i.e. $\mathbf{g_0} = (212)_p \to (012)_{gr}$ followed by $(012)_{gr} \to (212)_{ye}$, appears as a conventional twin relatively to the parent crystal because the plane $\mathbf{g_0} = (212)_p$ is restored, i.e. $(212)_p \to (212)_{ye}$.
4. The direction $\mathbf{u_0} = [0\bar{2}1]_p$ is maintained invariant by the three twinning modes, only its indexes are changed into equivalent ones: $[0\bar{2}1]_p \to [\bar{2}\bar{2}1]_{gr} \to [0\bar{2}1]_{ye}$

Now, let us define the crystallographic properties of the conventional (parent → yellow) twin. Its correspondence, distortion and orientation matrices are determined by combination of the matrices determined in the previous sections.

## Composition of the correspondence matrices

The first correspondence $\mathbf{C}_{hex}^{gr \to p}$ is followed by the second correspondence $\mathbf{C}_{hex}^{ye \to gr}$. Their composition is simply

$$\mathbf{C}_{hex}^{ye \to p} = \mathbf{C}_{hex}^{ye \to gr} \cdot \mathbf{C}_{hex}^{gr \to p} = \begin{pmatrix} \frac{1}{4} & \frac{5}{8} & \frac{5}{4} \\ 1 & -\frac{1}{2} & 1 \\ \frac{1}{4} & \frac{1}{8} & -\frac{3}{4} \end{pmatrix} \qquad (44)$$

$$\text{and } (\mathbf{C}_{hex}^{ye \to p})^* = \begin{pmatrix} \frac{1}{4} & 1 & \frac{1}{4} \\ \frac{5}{8} & -\frac{1}{2} & \frac{1}{8} \\ \frac{5}{4} & 1 & -\frac{3}{4} \end{pmatrix}$$

## Composition of the distortion matrices

The first distortion $\mathbf{D}_{hex}^{p \to gr}$ is followed by the second distortion $\mathbf{D}_{hex}^{gr \to ye}$. It is necessary to work in the same basis to compose these active matrices. The matrix $\mathbf{D}_{hex}^{gr \to ye}$ expressed in the parent hexagonal basis becomes $\mathbf{T}_{hex}^{p \to gr} \cdot \mathbf{D}_{hex}^{gr \to ye} \cdot \left(\mathbf{T}_{hex}^{p \to gr}\right)^{-1}$. The composition is thus

$$\mathbf{D}_{hex}^{p \to ye} = \mathbf{T}_{hex}^{p \to gr} \cdot \mathbf{D}_{hex}^{gr \to ye} \cdot \left(\mathbf{T}_{hex}^{p \to gr}\right)^{-1} \cdot \mathbf{D}_{hex}^{p \to gr} \tag{45}$$

$$= \begin{pmatrix} \frac{33}{28} - \frac{30}{7(3+7\gamma^2)} & \frac{5(-3+\gamma^2)}{8(3+7\gamma^2)} & \frac{5(-3+\gamma^2)}{4(3+7\gamma^2)} \\ \frac{-3+\gamma^2}{3+7\gamma^2} & \frac{3+15\gamma^2}{6+14\gamma^2} & \frac{-3+\gamma^2}{3+7\gamma^2} \\ -\frac{1}{4} + \frac{6}{3+7\gamma^2} & -\frac{1}{8} + \frac{3}{3+7\gamma^2} & \frac{3}{4} + \frac{6}{3+7\gamma^2} \end{pmatrix}$$

It is checked that $Det(\mathbf{D}_{hex}^{p \to ye}) = 1$, and that the directions $[0\bar{2}1]_p$ and $[\bar{1}01]_p$ are invariant by $\mathbf{D}_{hex}^{p \to ye}$ whatever the packing ratio. This proves that $\mathbf{D}_{hex}^{p \to ye}$ is a simple shear matrix that leaves invariant the plane $(212)_{hex}$. Consequently, $\mathbf{D}_{hex}^{p \to ye}$ is a *conventional* twinning matrix. The shear vector $\mathbf{s}$ is calculated by considering the normalized vector perpendicular to the plane $\mathbf{g} = (212)_p$ expressed in the orthonormal basis, i.e. $\mathbf{n} = \frac{\mathbf{H}_{hex}^* \mathbf{g}}{\|\mathbf{H}_{hex}^* \mathbf{g}\|}$. The shear vector is

$$\mathbf{s}_{ortho} = (\mathbf{D}_{ortho}^{p \to yellow} - \mathbf{I}) \cdot \mathbf{n} \tag{46}$$

When expressed in the hexagonal basis it becomes

$$\mathbf{s}_{hex} = (\mathbf{H}_{hex})^{-1} \cdot \mathbf{s}_{ortho} = \frac{3-\gamma^2}{\gamma\sqrt{9+21\gamma^2}} [\bar{5}, \bar{4}, 7]_{hex} \tag{47}$$

In four-index notation this vector is of type $\langle 1, 2, \bar{3}, 7\rangle_{hex}$. The shear amplitude is given by its norm, that can be calculated directly from $\mathbf{s}_{ortho}$. It is

$$s = \sqrt{\frac{7}{48}} \frac{|3-\gamma^2|}{\gamma} \tag{48}$$

In the case of hard-sphere packing ratio, the analytical expression of the distortion matrix takes rational values:

$$\mathbf{D}_{hex}^{p \to ye} = \begin{pmatrix} \frac{51}{52} & -\frac{1}{104} & -\frac{1}{52} \\ -\frac{1}{65} & \frac{129}{130} & -\frac{1}{65} \\ \frac{7}{260} & \frac{7}{520} & \frac{267}{260} \end{pmatrix} \approx \begin{pmatrix} 0.9808 & -0.0096 & -0.01922 \\ -0.0154 & 0.9923 & -0.0154 \\ 0.0269 & 0.0135 & 1.0269 \end{pmatrix} \tag{49}$$

The shear value along the direction $[\bar{5}, \bar{4}, 7]_{hex}$ is $s = \frac{1}{24}\sqrt{\frac{7}{2}} \approx 0.078$.

## Composition of the coordinate transformation matrices

The first coordinate transformation $\mathbf{T}_{hex}^{p \to gr}$ is followed by the second coordinate transformation $\mathbf{T}_{hex}^{gr \to ye}$. The composition of these passive matrices is simply

$$\mathbf{T}_{hex}^{p \to ye} = \mathbf{T}_{hex}^{p \to gr} \cdot \mathbf{T}_{hex}^{gr \to ye} = \begin{pmatrix} \dfrac{-3+3\gamma^2}{3+7\gamma^2} & \dfrac{5\gamma^2}{3+7\gamma^2} & \dfrac{10\gamma^2}{3+7\gamma^2} \\ \dfrac{8\gamma^2}{3+7\gamma^2} & -\dfrac{3(1+\gamma^2)}{3+7\gamma^2} & \dfrac{8\gamma^2}{3+7\gamma^2} \\ \dfrac{6}{3+7\gamma^2} & \dfrac{3}{3+7\gamma^2} & \dfrac{3-7\gamma^2}{3+7\gamma^2} \end{pmatrix} \quad (50)$$

An equivalent orientation matrix is obtained by using the internal symmetry $\mathbf{2}_{210}$; it is a rotation around the $\mathbf{u}_0 = [0\bar{2}1]_p$ axis and of angle

$$\theta_D = ArcCos\left(\dfrac{-6+11\gamma^2}{6+14\gamma^2}\right) \quad (51)$$

With the ideal hard sphere packing ratio, the disorientation matrix is rational

$$\mathbf{T}_{hex}^{p \to ye} \mathbf{2}_{210} = \begin{pmatrix} \dfrac{11}{13} & -\dfrac{8}{13} & -\dfrac{16}{13} \\ \dfrac{31}{65} & \dfrac{33}{65} & -\dfrac{64}{65} \\ \dfrac{27}{65} & -\dfrac{9}{65} & \dfrac{47}{65} \end{pmatrix} \quad (52)$$

And the rotation angle is $\theta_D = ArcCos\left(\dfrac{7}{13}\right) = 57.42°$

As the rotation between the parent and yellow twins $(p \to ye)$ is around the axis $\mathbf{u}_0 = [0\bar{2}1]_p$ and as this direction is also left invariant by the rotation associated with the $(p \to gr)$ twin and by the rotation associated with the $(gr \to ye)$ twin, it implies that the rotation angles should be linked by an addition. The rotation angles around the $\mathbf{u}_0$ axis are given in equations (28), (43) and (51), for the $(p \to gr)$, $(gr \to ye)$ and $(ye \to p)$ twins. Even if not obvious, it is indeed checked that

$$-ArcCos\left(\dfrac{6-\gamma^2}{2\sqrt{(3+\gamma^2)(3+7\gamma^2)}}\right) + ArcCos\left(\dfrac{-3-2\gamma^2}{\sqrt{(3+\gamma^2)(3+7\gamma^2)}}\right) = ArcCos(\dfrac{-6+11\gamma^2}{6+14\gamma^2}) \quad (53)$$

and thus also in the particular case of hard-sphere packing:

$$-ArcCos\left(\sqrt{\dfrac{5}{221}}\right) + ArcCos\left(-5\sqrt{\dfrac{5}{221}}\right) = ArcCos\left(\dfrac{7}{13}\right) \quad (54)$$

The disorientation, i.e. the equivalent rotation with the minimum rotation angle in absolute value, is obtained with the internal symmetry $\mathbf{2}_{110}$. The disorientation $\mathbf{T}_{hex}^{p \to ye} \mathbf{2}_{110}$ is a rotation around the axis $[\bar{2}21]_p$ and of angle

$$\theta_D = ArcCos\left(\dfrac{-6+13\gamma^2}{6+14\gamma^2}\right) \quad (55)$$

With the ideal hard sphere packing ratio, the disorientation matrix is rational

$$\mathbf{T}_{hex}^{p \to ye} \mathbf{2}_{110} = \begin{pmatrix} \frac{8}{13} & \frac{3}{13} & -\frac{16}{13} \\ -\frac{33}{65} & \frac{64}{65} & -\frac{64}{65} \\ \frac{9}{65} & \frac{18}{65} & \frac{47}{65} \end{pmatrix} \tag{56}$$

and the rotation angle is $\theta_D = ArcCos\left(\frac{43}{65}\right) = 48.58°$.

As the direction $[\bar{2}21]_p \equiv [241]_p$ by internal symmetry, the calculated disorientation (48.5°, $[\bar{2}21]_p$) between the parent crystal and the yellow twin fits exactly that obtained in the EBSD map (**Fig.1c**).

## 5. Summary of the calculations

In order to get a better overview of the results, a table summarizing the main crystallographic characteristics theoretically calculated is given below:

| Twin mode | Correspondence matrix | Distortion matrix | Misorientation matrix | Angle of rotation around the axis $u_0$ |
|---|---|---|---|---|
| $(p \to gr)$ | (17) | (23) | (25) | (28) |
| $(gr \to ye)$ | (35) | (40) | (42) | (43) |
| $(p \to ye)$ | (44) | (45) | (50) | (51) |

*Supplementary Table 2: Summary of the main crystallographic equations related to the three twin modes. The last column gives the rotation angle associated with the misorientation matrix chosen among the equivalent ones such that the rotation axis is **u₀**.*